\documentclass[aps,pre,preprint,groupedaddress]{revtex4-2}

\usepackage{fullpage}
\usepackage{todonotes,color}
\usepackage[centertags]{amsmath}
\usepackage{amscd,amsfonts,amssymb,latexsym}
\usepackage{graphicx}
\usepackage{theorem}
\usepackage{enumerate}
\usepackage[toc,page]{appendix}
\usepackage{float}
\usepackage{subcaption,overpic}
\usepackage{todonotes,color}

\newcommand{\la}{\lambda}
\newcommand{\non}{\nonumber}
\newcommand{\eps}{\epsilon}
\newcommand{\beqa}{\begin{eqnarray}}
\newcommand{\eeqa}{\end{eqnarray}}
\newcommand{\beqas}{\begin{eqnarray*}}
\newcommand{\eeqas}{\end{eqnarray*}}
\newcommand{\ba}{\begin{align}}
\newcommand{\ea}{\end{align}}
\newcommand{\bas}{\begin{align*}}
\newcommand{\eas}{\end{align*}}
\newcommand{\beq}{\begin{equation}}
\newcommand{\eeq}{\end{equation}}
\newcommand{\re}{\mathrm{Re}}
\newcommand{\im}{\mathrm{Im}}
\newcommand{\ra}{\rightarrow}
\newcommand{\al}{\alpha}

\newcommand{\ee}{\mathrm{e}}
\newcommand{\ii}{\mathrm{i}}
\newcommand{\pdhfrac}[2]{\mathchoice{\frac{#1}{#2}}{#1/#2}{#1/#2}{#1/#2}}
\newcommand{\fdd}[2]{\pdhfrac{\mathrm{d}#1}{\mathrm{d}#2}}
\newcommand{\sdd}[2]{\pdhfrac{\mathrm{d}^2#1}{\mathrm{d}#2^2}}
\newcommand{\pd}[2]{\pdhfrac{{\partial}#1}{{\partial}#2}}
\newcommand{\spd}[2]{\pdhfrac{\partial^2#1}{{\partial}#2^2}}

\renewcommand{\d}[1]{\mathrm{d}#1}

\newcommand{\Vs}{V_{s}}
\newcommand{\ps}{\Phi_{s}}
\newcommand{\laexp}{\lambda_{\mathrm{exp}}}
\newcommand{\fexp}{f_{\mathrm{exp}}}
\newcommand{\las}{\la_{\mathrm{reg}}}
\newcommand{\fs}{f_{\mathrm{reg}}}
\newcommand{\Asf}{{A^{\mathrm f}_{\mathrm{reg}}}}

\newcommand{\aF}{a_{\mathrm{f}}}
\newcommand{\bF}{b_{\mathrm{f}}}
\newcommand{\aG}{a_{\mathrm{g}}}
\newcommand{\Rf}{R_{\mathrm{f}}}
\newcommand{\Rg}{R_{\mathrm{g}}}
\newcommand{\bG}{b_{\mathrm{g}}}
\newcommand{\asf}{{a^{\mathrm f}_{\mathrm{reg}}}}

\newcommand{\asg}{{a^{\mathrm g}_{\mathrm{reg}}}}
\newcommand{\bsf}{{b^{\mathrm f}_{\mathrm{reg}}}}
\newcommand{\bsg}{{b^{\mathrm g}_{\mathrm{reg}}}}
\newcommand{\Bsf}{{B^{\mathrm f}_{\mathrm{reg}}}}

\newcommand{\fb}{f_b}
\newcommand{\gs}{g_{\mathrm{reg}}}

\newcommand{\gexp}{g_{\mathrm{exp}}}

\newcommand{\V}{V}

\DeclareMathOperator{\sign}{sign}

\graphicspath{{./}{./Figures/}}
\makeatletter
\def\input@path{{./}{./Figures/}}
\makeatother

\begin{document}

\title{A Spectral Analysis of the Nonlinear Schr{\"o}dinger Equation
in the Co-Exploding Frame}

\author{S. Jon Chapman}
\affiliation{Mathematical Institute, University of Oxford, AWB, ROQ, Woodstock Road, Oxford OX2 6GG}

\author{M. Kavousanakis}
\affiliation{School of Chemical Engineering, National Technical University of Athens, 15780, Athens, Greece}

\author{E.G. Charalampidis}
\affiliation{Mathematics Department, California Polytechnic State University, San Luis Obispo, CA 93407-0403, USA}

 \author{I.G. Kevrekidis}
\affiliation{Department of Chemical and Biomolecular Engineering \& \\
Department of Applied Mathematics and Statistics, Johns Hopkins University, Baltimore, MD 21218, USA}

\author{P.G. Kevrekidis}
\affiliation{Department of Mathematics and Statistics, University of
  Massachusetts, Amherst MA 01003-4515, USA}

\date{\today}

\begin{abstract}
The nonlinear Schr{\"o}dinger model is a prototypical
dispersive wave equation that features finite time blowup, 
either for supercritical exponents (for fixed dimension) or
for supercritical dimensions (for fixed nonlinearity exponent).
Upon identifying the self-similar solutions in the so-called
``co-exploding frame'', a dynamical systems analysis of their
stability is natural, yet is complicated by the mixed 
Hamiltonian-dissipative character of the relevant frame.
In the present work, we study the spectral picture of
the relevant linearized problem. We examine the 
point spectrum of 3 eigenvalue pairs associated with
translation, $U(1)$ and conformal invariances, as well
as the continuous spectrum. We find that two eigenvalues
become positive, yet are attributed to symmetries and 
are thus not associated with instabilities. In addition to a vanishing
eigenvalue, 3 more are found to be negative and real,
while the continuous spectrum is nearly vertical and on
the left-half (spectral) plane. Finally, the subtle 
effects of the boundaries are also assessed and their 
role in the observed weak eigenvalue oscillations is
clarified. 
\end{abstract}


\maketitle

\section{Introduction}

One of the central dispersive nonlinear partial 
differential equations of relevance to a wide
range of physical systems is the nonlinear 
Schr{\"o}dinger (NLS) model~\cite{ablowitz0,ablowitz01,ablowitz1,sulem,siambook}.
Among the different research themes where
the NLS plays a central role, one can mention the study of 
the electric field of
light in nonlinear optical systems~\cite{hasegawa,kivshar}, as well as in plasmas~\cite{plasmas},
the realm of water waves and the evolution of their height, e.g., in deep water~\cite{ir,mjarecent}, as well
as the condensate wavefunction for 
mean-field models of atomic Bose-Einstein 
condensates (BECs)~\cite{stringari,pethick,siambook}. 
The prototypical variants of the equation involve
the self-focusing~\cite{sulem,fibich0} and the 
self-defocusing nonlinearity~\cite{siambook}, and 
the respective dynamics revolve around bright~\cite{abdull}
and dark~\cite{djf} solitons.

In the case of self-focusing (self-attractive) nonlinearity,
and for sufficiently high dimension (for fixed nonlinearity)
or for sufficiently strong nonlinearity (for fixed dimension),
a key feature of the NLS model is the presence of collapse type
phenomena, that have also been explored in numerous 
books~\cite{sulem,fibich0,boyd}, as well as reviews~\cite{fibich,berge,pelin}. 
Indeed, the topic of finite time blow up of supercritical 
NLS solutions 
has been the objective of 
continued study both in the mathematical and in the physical literature; 
see, e.g., Refs.~\cite{pavel,pavel2,gadi} and~\cite{koch,sveta} (and also 
references therein) for only some recent examples. Importantly, the study 
of collapse is not only a mathematical idealization but rather has become 
accessible to physical experiments. In fact, on the one hand, there is 
the 
well-developed field of nonlinear optics, where not only the well-known, 
two-dimensional collapsing waveform of the Townes soliton has been 
observed~\cite{moll1} but also more elaborate themes have been touched 
upon including the collapse of optical vortices~\cite{vortex}, the loss 
of phase information of collapsing filaments~\cite{phase}, and the manipulation 
of the medium to avert optical collapse~\cite{psaltis}.
{On the other hand, a remarkable, very recent {experimental} development 
has been the emergence of {2 distinct} works in the atomic physics realm of BECs,
  observing Townes solitons in the $2d$ setting~\cite{bectownes,bectownes2}.
  Here, collapsing waveforms in higher dimensions
had been experimentally identified earlier~\cite{donley,cornish}, and the 
ability to manipulate the nonlinearity~\cite{haller} and the initial 
conditions~\cite{boris} has continued to  improve in recent times. 
In one of these recent works~\cite{bectownes} the modulational
instability was manipulated
to produce (in a less controllable, yet experimentally observable) way such
Townes waveforms.
The authors of the second work~\cite{bectownes2} leveraged a reduction of a minority component
in a two-component gas into a single-component one with {\it effectively attractive}
interactions to produce a collapsing Townes waveform.}

In many of the above mathematical works that study the
dynamics of collapse, both in dispersive systems
such as the NLS~\cite{sulem,fibich0}, but also even
in dissipative systems such as reaction-diffusion ones~\cite{galaktionov_book}, the emphasis is on identifying the
solution in a frame where it becomes steady, namely a self-similar
(or ``co-exploding'')
one~\cite{ren,budd,jfw,galak2,budd_recent}.
A similar approach is leveraged in dynamical systems
and partial differential equations (PDEs) when exploring
traveling waves which are identified as steady solutions
in a so-called co-traveling frame. In such settings, a
natural next step is to explore the spectral stability of the solutions in such a frame~\cite{bjorn,promislow}. However, in the realm
of the self-similar solutions, far fewer studies appear
to be exploring the spectral properties of the wave
in the co-exploding frame~\cite{wit1,wit2,siettos}. Indeed, in the context of NLS,
the only earlier approach to spectrally explore the collapse
problem concerns the earlier work of some of the authors~\cite{siettos}. In a recent work, we revisited
this topic, attempting to examine the self-focusing problem
as a bifurcation one, identifying its effective normal 
form~\cite{jon1}.
In the present study, we complement this approach by systematically
examining the spectrum of the self-similarly collapsing
solitary wave. 

Upon setting up the relevant linearization problem in the
self-similar frame (in Section~\ref{mathsetup}), 
our starting point will consist of observations of
the spectrum of the underlying
Hamiltonian system before the bifurcation point (in Section~\ref{numresults}). We will
examine the relevant spectral picture when approaching the
limit point where collapsing solutions emerge, and also
we will explore the same picture 
for the {\it dissipative} system that results in the
co-exploding frame past the critical point. In Section~\ref{theory},
having observed
the relevant spectrum, we will then turn to a more refined
analysis of the different eigenvalues thereof, one-by-one.
Finally, we will synthesize the picture and its dynamical
implications and offer some conclusions and future challenges
in Section~\ref{conclusions}. The Appendices offer some additional insights, 
including about how a symmetry of the original frame can turn
into an unstable eigendirection in a renormalized one, 
as well as about the role of the normal form obtained
previously in~\cite{jon1} in connection with the eigenvalues
identified herein.

\section{Basic Mathematical Setup}\label{mathsetup}

Our model of interest will be the one-dimensional,
general-nonlinearity-exponent
variant of NLS in the form
\begin{eqnarray}
\ii \pd{\psi}{z} + \spd{\psi}{x} + |\psi|^{2 \sigma} \psi = 0.
\label{nls1}
\end{eqnarray}
Notice that here we have used the typical optics notation,
where $z$ is the evolution variable, representing the propagation distance~\cite{sulem}.
This model has been studied extensively in ~\cite{sulem,fibich0}
and it is well-known that in $d$-dimensions, the condition for
its collapse is $\sigma d>2$. The model is subcritical
for $\sigma d<2$, and the special case of $\sigma d=2$ separates
the two regimes. We opt to consider the $d=1$ case for a
number of practical reasons, including (a) the availability
of an analytical solution for all values of
$\sigma$, namely $\psi=\ee^{\ii z} (1+\sigma)^{1/(2 \sigma)}
{\rm sech}\left[ (2 \sigma)^{1/2} (x-x_0) \right]$ and (b) the 
computational convenience of the relevant spectral
calculations. As we will see below, the latter will
be sensitively dependent on the domain size and
its boundary conditions, and associated considerations
will be even more delicate (and imposing  a substantial
additional computational overhead) in higher dimensions. 
Nevertheless, we expect the main features and techniques
proposed herein to be directly reflected in such 
higher-dimensional settings,
as will be evident in what follows. 

The Hamiltonian associated with Eq.~(\ref{nls1}) is given by
\begin{eqnarray} H = \int_{-\infty}^\infty \left(
\left| \pd{\psi}{x}\right|^2 -\frac{1}{\sigma+1} |\psi|^{2\sigma+2}
\right) \, \d x.
\label{ham2}
\end{eqnarray}
The dynamical equations satisfy:
\[ \ii \pd{\psi}{z} = \frac{\delta H}{\delta \psi^*}, \qquad \ii 
 \pd{\psi^*}{z} = -\frac{\delta H}{\delta \psi}.\]
We require that $H$ be finite.

In order to go to the co-exploding frame,
we introduce the well-known~\cite{sulem,fibich0} stretched variables,
rescaling space by the length scale $L(z)$
\begin{eqnarray}  \xi = \frac{x}{L(z)}, \quad \tau = \int_0^z
  \frac{\d z'}{L^2(z')}, \quad \psi(x,z) = L^{-1/\sigma}
  u(\xi,\tau), 
  \label{transf3}
  \end{eqnarray}
to give
\begin{eqnarray}
  \ii \pd{u}{\tau} + \spd{u}{\xi} +  |u|^{2 \sigma} u
  - \ii \xi L L_z \pd{u}{\xi}- \frac{\ii L L_z}{\sigma} u= 0,
  \label{pde4}
  \end{eqnarray}
  and the corresponding rescaling of the Hamiltonian:
  \begin{eqnarray} H = L^{-2/\sigma-2}\int_{-\infty}^\infty \left(
\left| \pd{u}{\xi}\right|^2 -\frac{1}{\sigma+1} |u|^{2\sigma+2}
\right) \, \d x.
\label{ham5}
\end{eqnarray}
We factor out the frequency of our solution without loss 
of generality and assign the rate of width shrinkage/amplitude
growth to be termed as $G$ by setting
\begin{eqnarray}
u(\xi,\tau) = \Phi(\xi,\tau) \ee^{\ii \tau}, \qquad G = -L L_z, 
\label{transf6}
\end{eqnarray}
which reduce Eq.~(\ref{pde4}) into
\begin{eqnarray}
  \ii \pd{\Phi}{\tau} +  \spd{\Phi}{\xi}+ |\Phi|^{2 \sigma}\Phi - \Phi 
+ \frac{\ii G}{ \sigma} \Phi
+ \ii G \xi \pd{\Phi}{\xi}= 0.
\label{pde7}
\end{eqnarray}

It is particularly important for our considerations that will
follow to emphasize that the above system bears a rather unusual
``mixed'' character. Along the manifold of $G=0$ (solitonic)
solutions, the relevant model falls back on the original one,
retaining its Hamiltonian structure. 
Nevertheless, for the genuinely self-similar solutions
of $G \neq 0$, the system is no longer conservative in nature.
Hence, we are dealing with a mixed Hamiltonian-dissipative system
and the dissipativity for $G \neq 0$ should be mirrored
in the spectrum of the self-similar solutions. This is contrary
to what is the case for the four-fold symmetric spectrum
of the $G=0$ solitons, for which if $\lambda$ is an 
eigenvalue, so are $-\lambda$, $\lambda^{*}$ and
$-\lambda^{*}$.

We will find it convenient to also perform an additional transformation 
by writing
\begin{eqnarray}
\Phi(\xi,\tau) = \V(\xi,\tau) \ee^{-\ii G(\tau) \xi^2/4}
\label{transf8}
\end{eqnarray}
to give
\begin{eqnarray}
  \ii \pd{\V}{\tau} + \frac{G' \xi^2}{4} \V+ \spd{\V}{\xi}+ |\V|^{2 \sigma}\V - \V 
- \frac{\ii (\sigma-2) G}{2 \sigma} \V
+ \frac{G^2 \xi^2 }{4}\V= 0,
\label{pde9}
\end{eqnarray}
where $G' = \fdd{G}{\tau}$, since then (without loss of generality) the imaginary part of $V$ is exponentially small in $G$ \cite{jon1}. Notice that above
we have suppressed the dependence of $G$ on the parameter $\sigma$.

Our principal aim as indicated above is to consider
the spectral stability of the steady-state solutions
(with $G \neq 0$) in the co-exploding frame. 
These correspond to self-similar blowup solutions in the original
frame. 
Such steady-state solutions denoted as $\ps$ satisfy
\begin{eqnarray}
  \sdd{\ps}{\xi}+ |\ps|^{2 \sigma}\ps - \ps 
+ \frac{\ii G}{ \sigma} \ps
+ \ii G \xi \fdd{\ps}{\xi}= 0,
\label{ode10}
\end{eqnarray}
or equivalently
\beq
 \sdd{\Vs}{\xi}+ |\Vs|^{2 \sigma}\Vs - \Vs 
- \frac{\ii (\sigma-2) G}{2 \sigma} \Vs
+ \frac{G^2 \xi^2 }{4}\Vs= 0.\label{sseqn}
\eeq
%
We now linearize Eq.~(\ref{pde7}) about the relevant
steady-state solutions in the co-exploding frame, $\ps$, by setting:
\begin{eqnarray}
 \phi(\xi,\tau) = \ps(\xi) + \eps\left( X(\xi) \ee^{\la \tau} + Y^*(\xi) \ee^{\la^* \tau} \right),
 \label{linear13}
 \end{eqnarray}
giving rise to the operator eigenvalue problem
\beqa
\ii\la X & = & \left(-\sdd{ }{\xi}   
- ( \sigma+1)|\ps|^{2 \sigma}   
+ 1
- \frac{\ii  G}{ \sigma} 
- \ii G \xi \fdd{}{\xi}\right)X
-\sigma |\ps|^{2\sigma-2}\ps^{2} Y,\label{x-eqn}\\
\ii  \la Y & = & 
  \sigma |\ps|^{2\sigma-2}(\ps^*)^{2} X +
\left( \sdd{}{\xi} 
+ ( \sigma+1)|\ps|^{2 \sigma} 
-  1
- \frac{\ii G}{\sigma}  
- \ii G \xi \fdd{}{\xi}\right)Y.\label{y-eqn}
\eeqa
In a similar vein, for  the stationary solution $V_s(\xi)$,
we linearize in $\eps$ by writing
\begin{eqnarray}
\V(\xi,\tau) = \Vs(\xi) + \eps\left( f(\xi) \ee^{\la \tau} + g^*(\xi) \ee^{\la^* \tau} \right)
\label{linear14}
\end{eqnarray} 
which  leads to
\beqa
\ii\la f+\sdd{ f}{\xi} +
\sigma |\Vs|^{2\sigma-2}\Vs^{2} g  
+ ( \sigma+1)|\Vs|^{2 \sigma}  f 
- f
- \frac{\ii (\sigma-2) G}{2 \sigma} f
+ \frac{G^2 \xi^2 }{4} f &=& 0,\label{geqn}\\
-\ii  \la g
+ \sdd{g}{\xi} 
+  \sigma |\Vs|^{2\sigma-2}(\Vs^*)^{2} f 
+ ( \sigma+1)|\Vs|^{2 \sigma} g 
-  g 
+ \frac{\ii (\sigma-2) G}{2 \sigma}  g 
+ \frac{G^2 \xi^2 }{4} g  &=& 0.\label{feqn}
\eeqa
On the finite  domain $[-K,K]$ we impose the boundary conditions
\[ \pd{\phi}{\xi} = 0 \qquad \mbox{ at } \xi = \pm K,\]
which corresponds to
\[ \pd{V}{\xi} = \pm \frac{\ii G K V}{2} \qquad \mbox{ at } \xi = \pm K.\]
For the perturbation this gives, correspondingly, 
\[ \pd{X}{\xi} = \pd{Y}{\xi} = 0,  \qquad  \pd{f}{\xi} = \pm \frac{\ii G K f}{2}, \qquad \pd{g}{\xi} = \mp \frac{\ii G K g}{2} \qquad \mbox{ on } \xi = \pm K.\]
Let us now try to explore, on the basis of the above
principal setup, what we should expect to see in the 
linearization around a collapsing waveform.

\section{Principal Numerical Results}\label{numresults}

The question of how the spectrum changes under the
type of nontrivial scaling transformation discussed above
requires particular attention. This topic was first addressed
systematically, to the best of our knowledge, in a different
class of systems, in the pioneering work
of~\cite{wit1,wit2} who realized that such a transformation 
that rescales space and time may have  profound
implications within the renormalized frame 
as regards the interpretations of symmetries
of the original frame.
To explain this subtle point, we provide arguably the
simplest possible example that we have been able to identify
in  Appendix A of the present manuscript. There, and in 
the cleaner/simpler setting of an autonomous ordinary differential
equation, it can be seen that the symmetry of time translation
of the original system leads to an ``apparent instability''
in the renormalized frame. This is because a shift in, e.g., the
time of collapse in the original frame, due to the exponential
nature of the transformation between the renormalized and the regular
time, leads to an exponential deviation in the renormalized
frame and hence an {\it apparent} instability. 

The key take-home message from this example is that {\it symmetries
of the original frame may no longer correspond to ones
such in the renormalized frame}. The even more dire consequence
is that {\it symmetries of the original frame may appear
  as instabilities in the renormalized one.}
For example, differentiating Eq.~(\ref{ode10}) with respect to $\xi$ gives
\begin{eqnarray}
   \frac{\d^3\ps}{\d\xi^3}+ (\sigma+1)|\ps|^{2\sigma}\fdd{\ps}{\xi} + \sigma |\ps|^{2\sigma-2}\ps^{2}\fdd{\ps^*}{\xi} - \fdd{\ps}{\xi}
+ \frac{\ii G}{ \sigma} \fdd{\ps}{\xi}
+ \ii G \xi \sdd{\ps}{\xi}+ \ii G  \fdd{\ps}{\xi}= 0,\qquad
\label{ode11}
\end{eqnarray}
from which we  observe  that $X=\fdd{\phi_s}{\xi}$
and $Y=\fdd{\phi_s^{*}}{\xi}$ satisfy Eqs.~(\ref{x-eqn})-(\ref{y-eqn}) if
we choose $\lambda=G$.
The eigenvector is
associated with the derivative, which is well-known to be
the generator of translations. However, instead of this vector being
associated with a neutral direction, it is now associated
with an ``apparently unstable'' eigenmode (since $G>0$). Nevertheless,
that eigenmode is {\it not} a true instability in the original
frame, even though it appears as one in the renormalized frame.
Rather, it only involves spatial translation, i.e., a symmetry,
and its 
suitable reinterpretation in this renormalized frame.

Armed with this important piece of understanding,
let us now scrutinize the spectral picture in further detail.
As is natural, we start with the subcritical case
of $\sigma d<2$. In the integrable limit of $d=\sigma=1$,
it is well-known~\cite{kaup} that the spectrum of the linearization
 of the NLS soliton possesses two neutral directions,
one associated with spatial translations, and one
associated with the phase or gauge (U$(1)$) invariance. We already saw that the
derivative $\fdd{\phi_s}{\xi}$ is connected to the translational
eigenvector while the solution ($\phi_s$) itself is associated
with the corresponding phase eigenvector. In each case,
the generalized eigenvectors are known as well~\cite{kaup}.

\begin{figure}
  \begin{center}
  \begin{overpic}[width=0.95\textwidth]{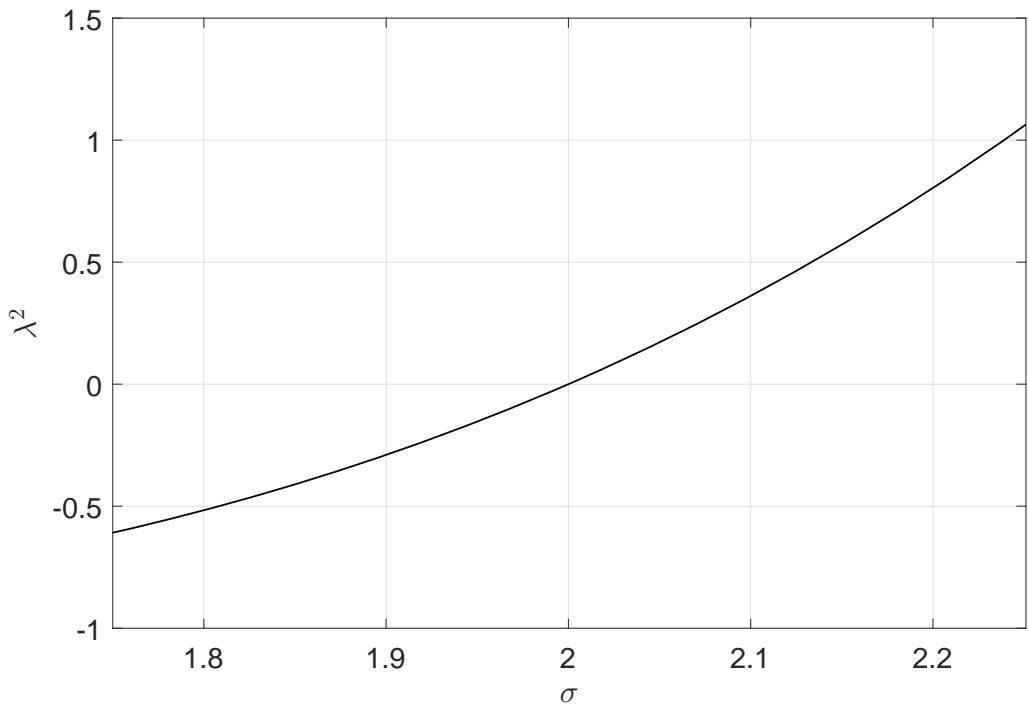}
\end{overpic}
    \end{center}
  \caption{Square of eigenvalue bifurcating from the band edge of the continuous spectrum, tending towards the origin as $\sigma\rightarrow2$, and finally giving rise to real eigenvalues (one positive and one negative) past the critical point, $\sigma=2$.}
  \label{fig:realpairvariation}
\end{figure}

\begin{figure}
  \begin{center}
  \begin{overpic}[width=0.95\textwidth]{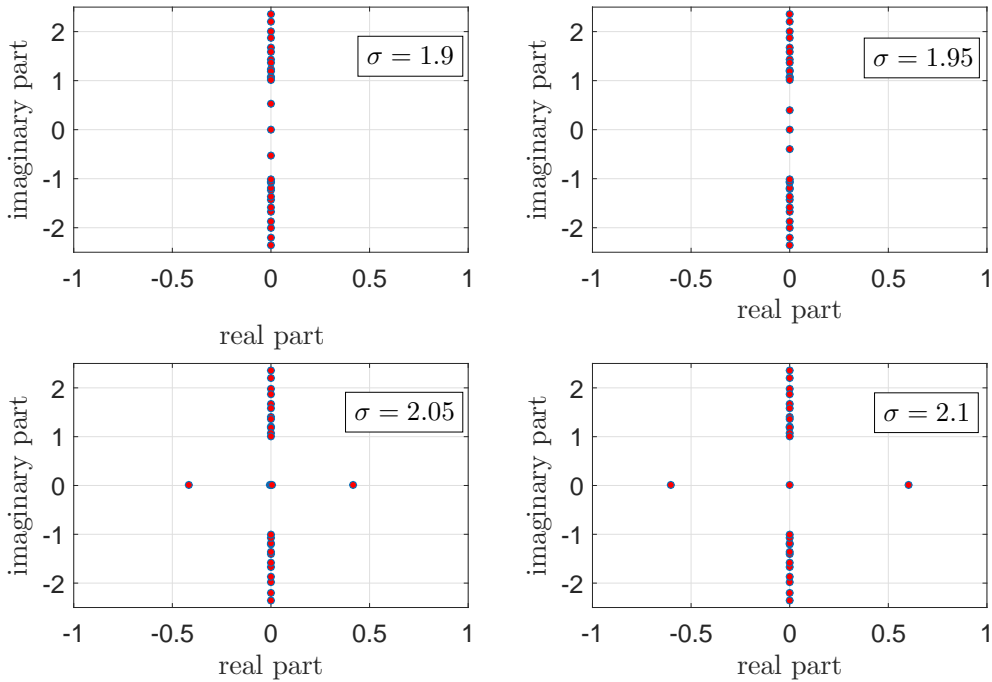}
\end{overpic}
    \end{center}
  \caption{Spectra of the numerically obtained soliton solution of the rescaled NLS equation with $K=20$ and $\sigma=1.9$, $\sigma=1.95$, $\sigma=2.05$ and $\sigma=2.1$.}
  \label{fig:solitonspectracomparison}
\end{figure}

As we depart (parametrically in $\sigma$) from the integrable limit, an eigenvalue pair
bifurcates from the band edge of the continuous spectrum
which consists of the union of the intervals
$\ii [1, \infty)$ and $-\ii [1,\infty)$ and tends (along
the imaginary axis) towards the origin
as $\sigma \rightarrow 2$. It is this eigenvalue pair that arrives
at the origin of the spectral plane, precisely at $\sigma=2$,
instituting the conformal invariance of the model, i.e., 
the invariance with respect to rescaling that paves the
way to collapse dynamics. The dependence
of this eigenvalue on the parameter $\sigma$ is 
shown in Fig.~\ref{fig:realpairvariation}. 
The corresponding eigenvector
in this case is
\begin{eqnarray}
X(\xi)=\ii \phi_s + G \left(\frac{\phi_s}{\sigma} + \xi \fdd{\phi_s}{\xi} \right),
\label{eigv1}
\end{eqnarray}
and similarly its conjugate yields $Y(\xi)$.
Past the critical point, the relevant eigenvalue becomes
real, giving rise to the dynamical instability of the soliton
and the emergence of the collapsing branch of solutions.
The spectra of the solitonic solution of $G=0$ for 
$\sigma$ below the critical one (of $\sigma_c=2$ for $d=1$)
and above the critical point are shown 
in Fig.~\ref{fig:solitonspectracomparison}.

As the bifurcation of the self-similarly focusing branch
of solutions occurs~\cite{siettos,jon1}, 
the natural question is what becomes
of the spectrum and what are the corresponding dynamical
implications of this spectral linearization picture.
Recall that at the critical point, the ``parent branch'' of
solitary waves has, in addition to the above mentioned 
continuous spectrum, 3 eigenvalue pairs at the origin. 
Hence, as this Hamiltonian system turns dissipative for
$G >0$, we have to determine the fate of the 6 eigenvalues
stemming from the origin, and the associated continuous
spectrum band. Notice that the 6 eigenvalues will
{\it no longer} constitute pairs, except perhaps approximately,
as the dissipativity of $G \neq 0$ 
destroys the Hamiltonian character and hence the eigenvalue pairing.

\begin{figure}
  \begin{center}
  
\begin{overpic}[width=0.95\textwidth]{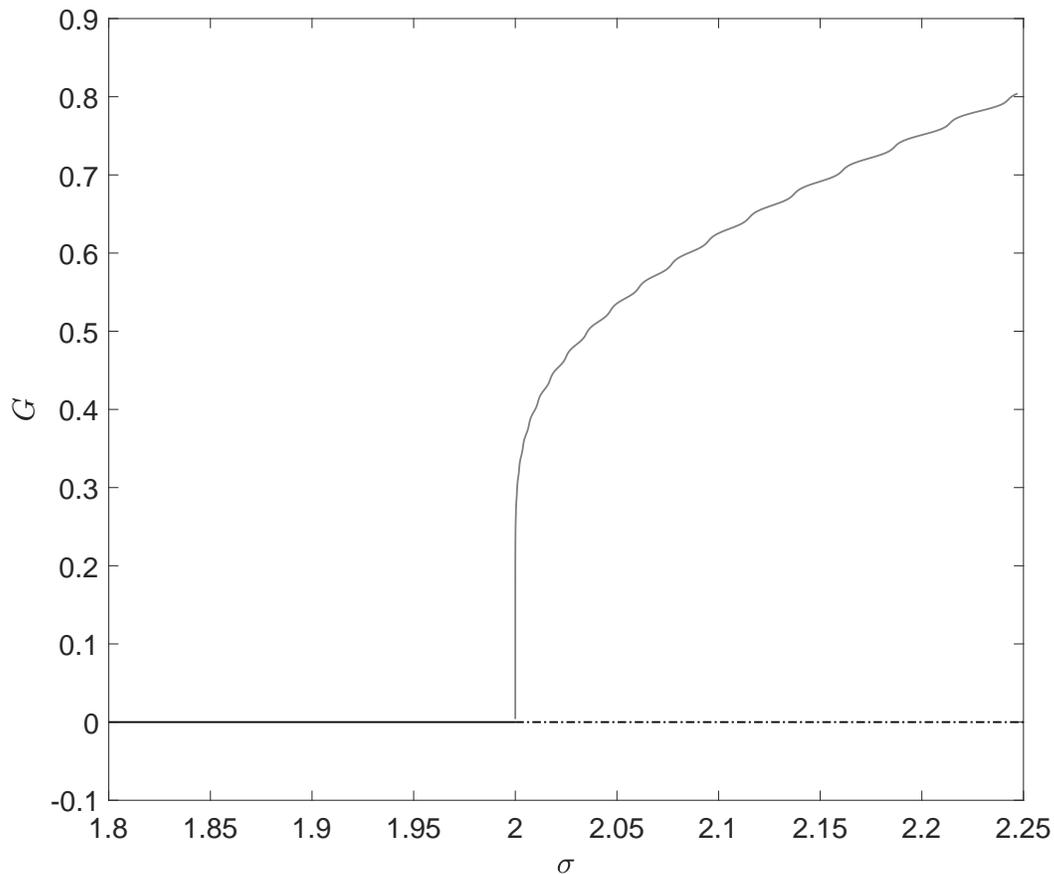}
\end{overpic}
    \end{center}
  \caption{Variation of blowup rate $G$ as a function of $\sigma$, for domain size $K=20$. The solitonic branch ($G=0$) remains stable up to $\sigma=2$ (black solid line) and becomes unstable for $\sigma>2$ (black dash-dotted line). The stable collapsing branch ($G>0$) is illustrated with solid grey line.
  }
  \label{fig:bifdiagram}
\end{figure}

\begin{figure}
\begin{center}
\begin{overpic}[width=0.95\textwidth]{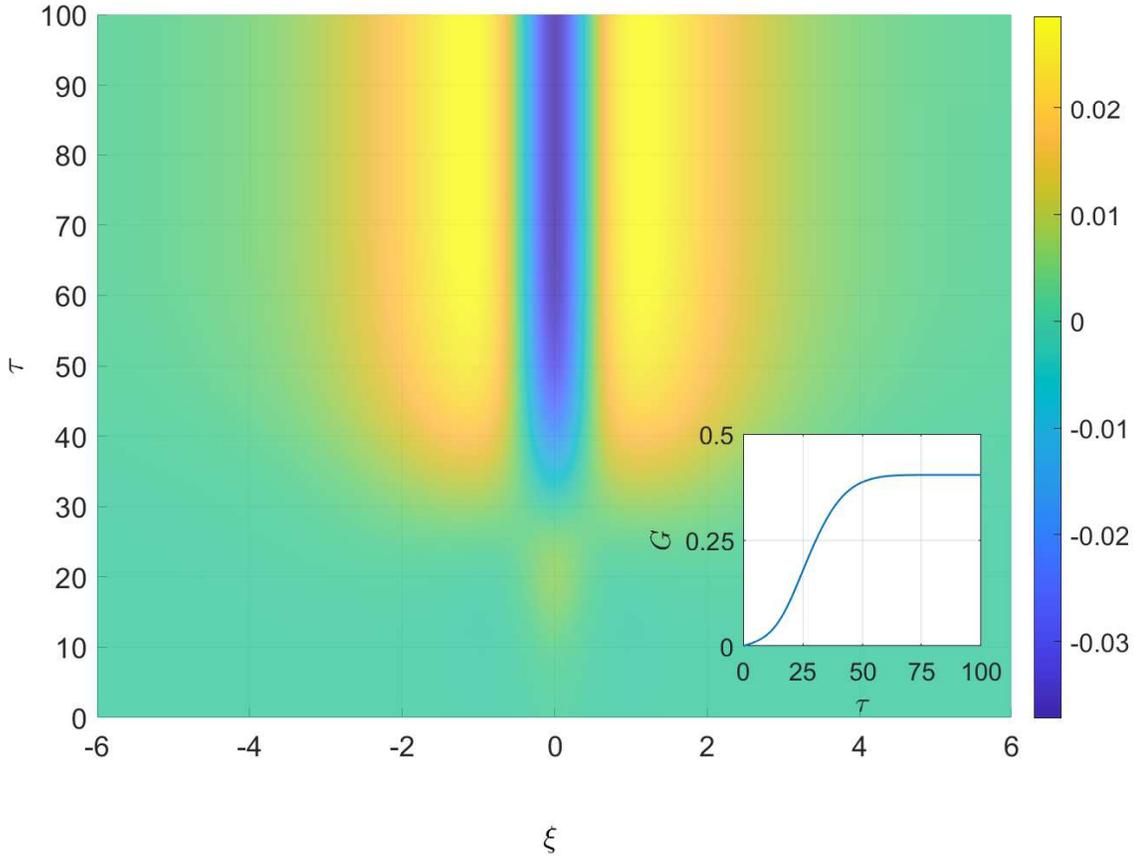}
\end{overpic}
    \end{center}
  \caption{Dynamics of $|\phi(\xi,\tau)|^2-|\phi(\xi,0)|^2$ in the co-exploding frame (rescaled NLS). The initial condition, $\phi(\xi,0)$ is the soliton solution for $\sigma=2.019$. Upon perturbing $\sigma$ to $\sigma=2.02$, the co-exploding dynamics converges to a ``steady-state'' solution. The inset on the bottom right illustrates the evolution of the blowup rate, $G$, with the rescaled time, $\tau$.}
  \label{fig:nls_dynamics}
\end{figure}

\begin{figure}
  \begin{center}
  
\begin{overpic}[width=0.95\textwidth]{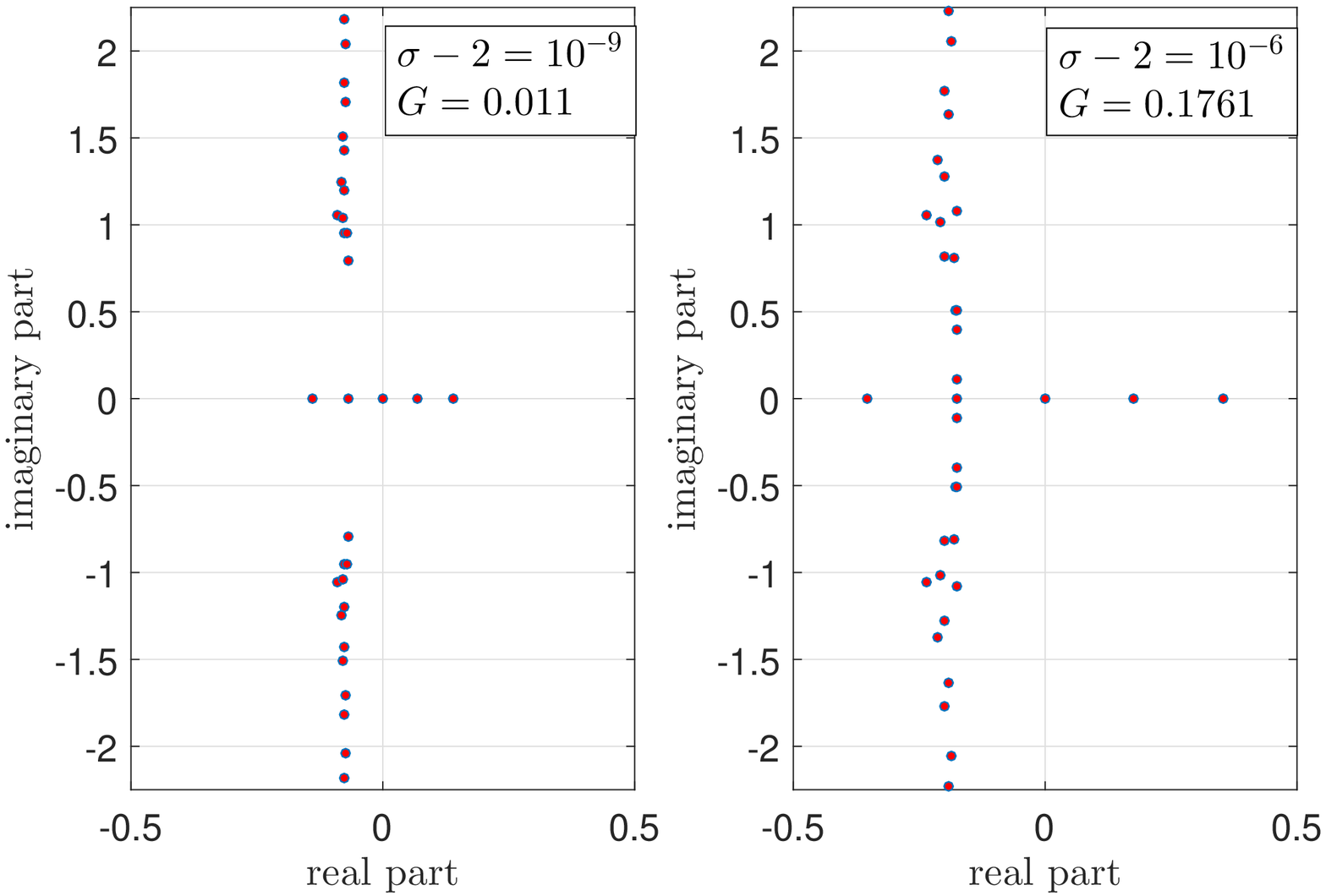}
\end{overpic}
\end{center}
\begin{center}
\begin{overpic}[width=0.95\textwidth]{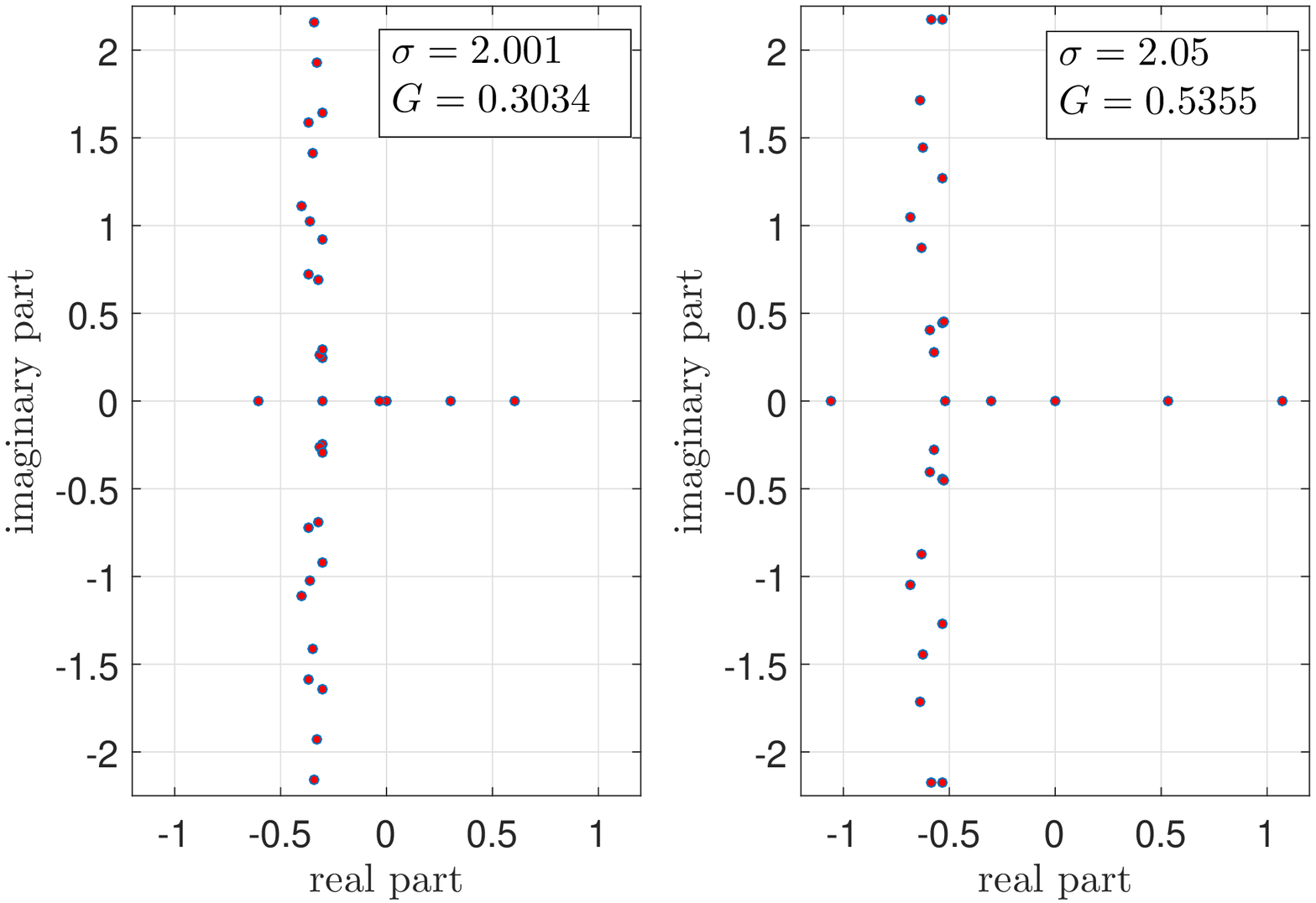}
\end{overpic}
    \end{center}
  \caption{Spectra obtained from the numerical solution of the rescaled NLS equation with $K=20$ and $\sigma$ values close to the critical value, $\sigma=2$: $\sigma=2+10^{-9}$ (top left panel) and $\sigma=2+10^{-6}$ (top right panel), as well as $\sigma=2.001$ (bottom left panel), $\sigma=2.05$ (bottom right panel).}
  \label{fig1}
\end{figure}

Before we systematically answer this question, it is relevant 
to remind the reader of the established bifurcation of 
solutions with nontrivial $G \neq 0$ for $\sigma>2$~\cite{sulem,siettos,jon1}.
The bifurcation diagram of the relevant solutions with a finite blowup rate $G$ 
is shown in Fig.~\ref{fig:bifdiagram}, and a typical example
of the associated waveforms and the dynamics of approaching
them within the realm of the mixed Hamiltonian-dissipative system 
of Eq.~(\ref{pde7}) is shown in Fig.~\ref{fig:nls_dynamics}.
The latter suggests the attractivity of the structures
and hence predisposes us towards their (effective) spectral stability.
Having obtained these solutions with finite non-vanishing $G$
as stationary ones (see the details in~\cite{jon1}), we are
now ready to solve the corresponding spectral problem for the
eigenvalues $\lambda$ and eigenvectors $(X,Y)$. Some typical examples
of the spectral plane of the imaginary vs. the real part of the
eigenvalues 
for specific choices of $\sigma$ (and hence $G$, per Fig.~\ref{fig:bifdiagram})
are shown in Fig.~\ref{fig1}.

\begin{figure}
  \begin{center}
\begin{overpic}[width=0.95\textwidth]{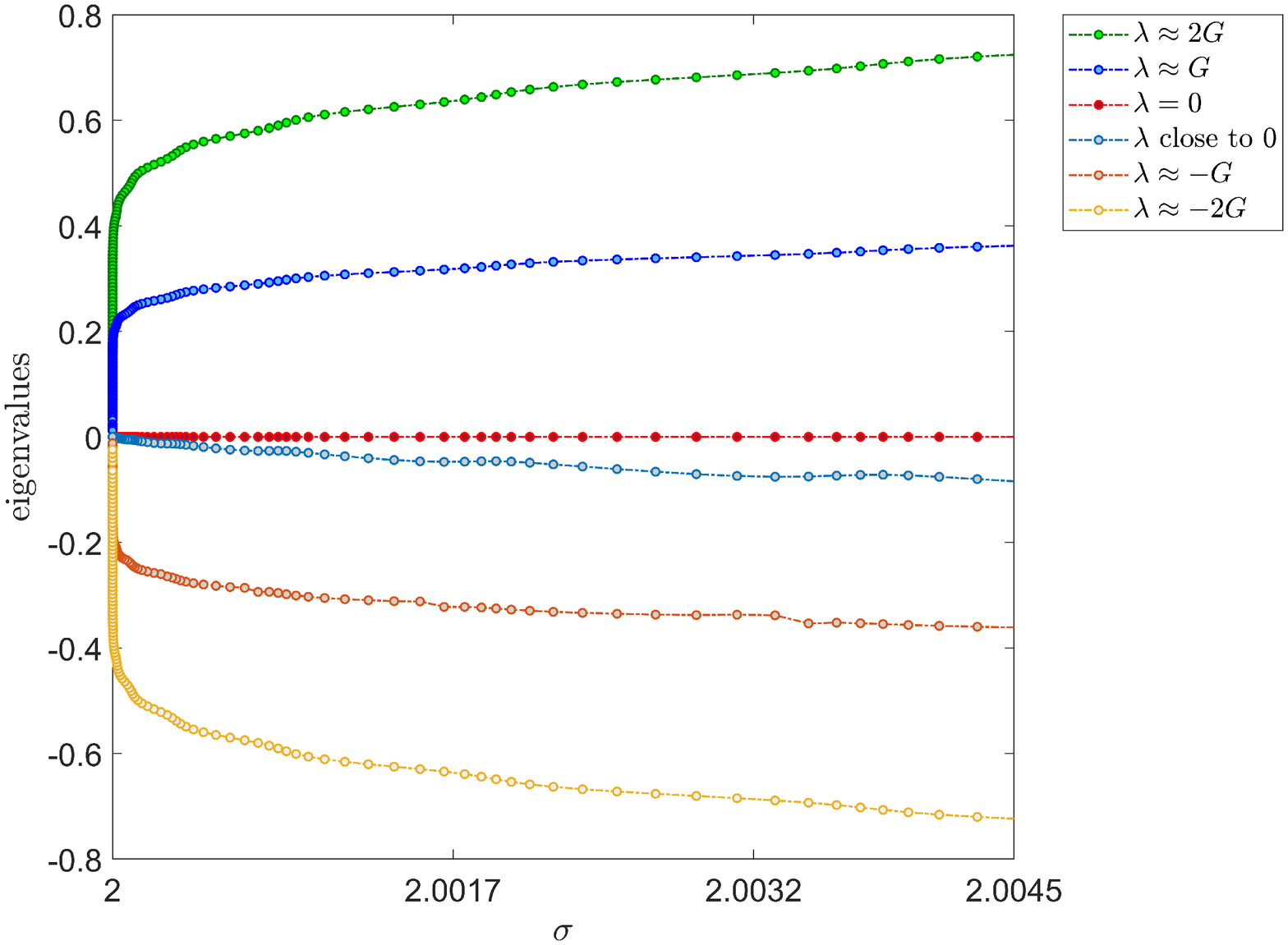}
\end{overpic}
    \end{center}
  \caption{Variation of $\lambda \approx 2G$, $\lambda \approx G$, $\lambda = 0$, $\lambda \approx -G$, $\lambda \approx -2G$ and $\lambda$ close to 0 eigenvalues with $\sigma$ values as obtained from the numerical solution of the rescaled NLS equation with $K=20$.
    }
  \label{fig2} 
\end{figure}

The answer to this central question of our manuscript for the
spectrum at the co-exploding frame is given in Fig.~\ref{fig2}.
There we can see that, in fact, {\it only one} out of the 6 eigenvalues stays at the origin. Indeed, $X \propto \phi_s$
remains an eigenvector with vanishing eigenvalue, as the 
rescaled model retains the original phase invariance.
Nevertheless, as indicated above, the generalized eigenvector
is no longer there (due to dissipativity) and, thus, the
associated eigenvalue acquires a small negative value. In addition, there are two pairs of
eigenvalues that are only {\it nearly} symmetric. We
find these to be at $\lambda \approx \pm 2 G$ and
$\lambda \approx \pm G$. All of these point spectrum
eigenvalues are systematically captured in Fig.~\ref{fig2}
to which we will return shortly. Moreover, there are two
more observations in place regarding Fig.~\ref{fig1}. One
of the above 6 eigenvalues (and one of the ones shown in Fig.~\ref{fig2},
as well), the eigenvalue at $\lambda \approx -G$, is hard to detect.
This is because it almost coincides with a nearly vertical line
of continuous spectrum with real part $\lambda_r = - G$, i.e., 
the continuous spectrum is approximately $\lambda = -G + \ii s$ for arbitrary
real $s$ (see also Appendix~\ref{sec:large}). 

As we already discussed above, the pair at $\lambda \approx \pm G$
is associated with spatial translation. Indeed, the derivative
eigenvector, through the exact calculation above, yields an eigenvalue of $\lambda = G$ in the infinite 
domain. It can be discerned from Fig.~\ref{fig2}, that this eigenvalue
is no longer {\it exactly} at $G$ on the finite domain but rather presents slight
undulations in its dependence. Indeed, one of our key aims
in the detailed calculations that follow will be to capture
these  finite-domain-induced undulatory corrections.
On the other hand, the eigenvalue at $-G$ is no longer exact
{\it even in the infinite domain}, due to the lack of symmetry,
as induced by the dissipative terms $\propto G$ in our linearized 
equation for $(X,Y)$ (or for $(f,g)$). In a very similar tenor,
the eigenvalue $\lambda=2 G$ is also exact in the infinite
domain limit, as can be verified by direct calculation, upon 
substituting the eigenvector
of Eq.~(\ref{eigv1}) in the linearization equations.
However, in this case too, the finite domain correction (to be
also evaluated below) induces an undulatory dependence on top
of the $\lambda=2 G$ leading order. Furthermore,  the eigenvalue $\la = -2 G$
is also no longer exact even for an infinite domain (due to dissipativity)
and in addition, there is an undulation (from the finite domain)
on top of this eigenvalue as well. This summary then accounts
for all the point spectrum eigenvalues.

It is relevant to add here two important observations.
The first one concerns the dynamics of the collapsing 
solutions. On the one hand, we obtain that the relevant
waveforms have two unstable eigendirections in the co-exploding
frame. However, on the other hand, we have illustrated
through our explicit calculations above (see also the pertinent
Appendix~A) that such eigendirections do not pertain to true
instabilities, but rather to neutral directions of the original
frame (spatial translations and rescalings of the original solution).
Given the rescaling of space and time in the co-exploding
frame, both of these actions move solutions exponentially far from other members of the family of such equivariant solutions, and thus appear as instabilities  in the co-exploding
frame, yet this is {\it not a true instability} in the original frame.
Hence, in line with our above dynamical evolution results, 
we expect such collapsing solutions to be dynamically
robust (modulo symmetries).

The second observation is related to the results for the spectrum given
in the earlier work of~\cite{siettos}. There, only one of these
positive eigenvalues was found and moreover the continuous spectrum
had a wider apparent extent around $\lambda_r \approx -G$ (extending
to values with more negative real part). The former of these
features was because the calculation of~\cite{siettos} was done
in the half-domain and hence, e.g., spatial translations were
a priori excluded from consideration. Furthermore, we believe
that the observations of the continuous spectrum had to do with
the discretization used in the latter case. Our refined numerics
here suggest that the continuous spectrum progressively tends
to the vertical line with $\lambda_r=-G$ (asymptotically
for large imaginary part). Finally, we also note that the main features of the computed spectra show only slight changes by increasing the size of the computational domain (see Appendix \ref{sec:Keffect}). 
Admittedly, in what
follows we can only offer an asymptotic prediction for the part
of the spectral band
with sufficiently large imaginary part. For the part with small
imaginary part, the situation is rather complex and constitutes
a technical challenge for potential future studies. Nevertheless,
we believe that we hereby offer a far more definitive 
perspective of both the point and the continuous spectrum, than was
previously available.

In the analytical calculations that follow (and their comparison
with the detailed numerical computations as regards the eigenvalue
corrections), we will consider each of these eigenvalues one by one.
We will split their dependence into a principal part (that we have
effectively already discussed above), and a correction that
stems either from the finiteness of the computational domain
(in the case of $\lambda=G$ or $\lambda=2G$) or from both
the inexactness of the symmetry in the dissipative system
{\it and} the finiteness of the computational domain
(in the case of the negative point spectrum eigenvalues). 
We will develop a solvability based approach to calculate
the residual of each of these eigenvalues and will subsequently
compare it to our systematic eigenvalue computations. Finally,
we will corroborate our theoretical conclusion on the 
effective spectral stability (modulo the symmetries) of the
collapsing solutions via direct numerical simulations in both
the original and the co-exploding frame.

For the performance of numerical computations, we adopt a fourth-order central finite difference scheme for the approximation of spatial derivatives. Space, $\xi \in [-K,K]$ is uniformly discretized with step, $d\xi=0.01$. Time integration (where needed) is performed utilizing MATLAB's ode23t ODE solver. Steady-state solutions are obtained through the iterative Newton-Raphson algorithm. Finally, the eigenvalue computations were performed by utilizing MATLAB's eig solver and corroborated further by using the contour-integral based FEAST eigenvalue solver \cite{kestyn_eric_tang} (and references therein). The spectral stability analysis results we obtained through the use of both eigenvalue solvers match precisely with each other. 

\section{Theoretical Analysis Approach}
\label{theory}
For our theoretical analysis, we work in terms of $\Vs$, $f$ and $g$.
We first outline the general methodology, before we apply it to each eigenvalue of the discrete spectrum in turn.

Suppose we have an asymptotic approximation to the eigenfunctions $\fs$ and $\gs$ and eigenvalue $\las$ which is accurate to all orders in $G$ but misses exponentially small terms. [Notice that in what follows, for mathematical convenience,
we will generally expand in powers of
$G$, rather than the parameter $\sigma$.]
Let us write $\la  =\las + \laexp$.
Then, Eqs.~(\ref{geqn})-(\ref{feqn}) give:
\beqas
\ii \las f+\sdd{ f}{\xi} +
\sigma |\Vs|^{2\sigma-2}\Vs^{2} g  
+ ( \sigma+1)|\Vs|^{2 \sigma}  f 
- f
- \frac{\ii (\sigma-2) G}{2 \sigma} f
+ \frac{G^2 \xi^2 }{4} f &=& -\ii \laexp f,\\
- \ii \las g
+ \sdd{g}{\xi} 
+  \sigma |\Vs|^{2\sigma-2}(\Vs^*)^{2} f 
+ ( \sigma+1)|\Vs|^{2 \sigma} g 
-  g 
+ \frac{\ii (\sigma-2) G}{2 \sigma}  g 
+ \frac{G^2 \xi^2 }{4} g  &=& \ii \laexp g.
\eeqas
If we multiply by $\fs$ and  $\gs$ respectively, add and integrate by 
parts, the left-hand side is
\beqas
&&\hspace{-0.5cm}\int_{-K}^{K}\left(\ii \las f+\sdd{ f}{\xi} +
\sigma |\Vs|^{2\sigma-2}\Vs^{2} g  
+ ( \sigma+1)|\Vs|^{2 \sigma}  f 
- f
- \frac{\ii (\sigma-2) G}{2 \sigma} f
+ \frac{G^2 \xi^2 }{4} f \right) \fs\, \d \xi \\
&& \hspace{-0cm}+
\int_{-K}^{K}\left(- \ii \las g
+ \sdd{g}{\xi} 
+  \sigma |\Vs|^{2\sigma-2}(\Vs^*)^{2} f 
+ ( \sigma+1)|\Vs|^{2 \sigma} g 
-  g 
+ \frac{\ii (\sigma-2) G}{2 \sigma}  g 
+ \frac{G^2 \xi^2 }{4} g\right) \gs\, \d \xi\\
& = & \int_{-K}^{K}\left(\ii \las  \fs+\sdd{ \fs}{\xi} 
+(\sigma+1)|\Vs|^{2\sigma}  \fs 
- \fs
- \frac{\ii(\sigma-2)G}{2 \sigma} \fs
+ \frac{G^2 \xi^2 }{4} \fs\right) f\, \d \xi\\
&& \int_{-K}^{K}\left(-\ii \las    \gs +\sdd{\gs}{\xi} 
+ (\sigma+1)|\Vs|^{2\sigma} \gs 
-  \gs + \frac{\ii(\sigma-2)G}{2 \sigma} \gs
+ \frac{G^2 \xi^2 }{4} \gs\right) g\, \d \xi
\\
&& \mbox{ }+\int_{-K}^{K}
\sigma |\Vs|^{2 \sigma-2}\Vs^2 g \fs +  \sigma |\Vs|^{2\sigma-2}(\Vs^*)^2 f \gs\, \d \xi + \left[ \fs \fdd{f}{\xi} - f \fdd{\fs}{\xi} +  \gs \fdd{g}{\xi} - g \fdd{\gs}{\xi}\right]^K_{-K}\\
& = & \int_{-K}^{K}
\Rf
 f +\Rg
 g + \sigma|\Vs|^{2 \sigma-2}(\gs f-\fs g)((\Vs^*)^2-\Vs^2)\, \d \xi \\ && \mbox{ }+ \left[ \fs \fdd{f}{\xi} - f \fdd{\fs}{\xi} +  \gs \fdd{g}{\xi} - g \fdd{\gs}{\xi}\right]^K_{-K},
 \eeqas
 where
 \beqas
 \Rf & = & \ii \las \fs+\sdd{ \fs}{\xi} 
+(\sigma+1)|\Vs|^{2\sigma}  \fs +  \sigma |\Vs|^{2\sigma-2}\Vs^2  \gs
- \fs
- \frac{\ii(\sigma-2)G}{2 \sigma} \fs
+ \frac{G^2 \xi^2 }{4} \fs,\\
\Rg & = &-\ii \las   \gs +\sdd{\gs}{\xi} 
+ (\sigma+1)|\Vs|^{2\sigma} \gs  +
\sigma |\Vs|^{2 \sigma-2}(\Vs^*)^2   \fs
-  \gs + \frac{\ii(\sigma-2)G}{2 \sigma} \gs
+ \frac{G^2 \xi^2 }{4} \gs,
\eeqas
are the exponentially small residuals from the regular asymptotic expansion.
Since the imaginary part of $\Vs$ is exponentially small, and $\fs$ and $\gs$ are exponentially close to $f$ and $g$, the term $(\gs f-\fs g)((\Vs^*)^2-\Vs^2)$ is doubly exponentially small and can be neglected.
Then, evaluating also the right-hand side, 
\begin{multline}
\int_{-K}^{K}
(\Rf f +\Rg g)\, \d \xi+ \left[ \fs \fdd{f}{\xi} - f \fdd{\fs}{\xi} +  \gs \fdd{g}{\xi} - g \fdd{\gs}{\xi}\right]^K_{-K}   \sim \ii \laexp\int_{-K}^K (\gs g- \fs f )\, \d \xi.\label{eig}
\end{multline}
Since $\Rf$, $\Rg$ and $\laexp$ are already exponentially small, we can use $\fs$, $\gs$ in place of $f$ and $g$ except in the boundary terms, introducing only double-exponentially-small errors. Thus, to exponential accuracy, 
\begin{multline}
\int_{-K}^{K}
(\Rf \fs +\Rg \gs)\, \d \xi+ \left[ \fs \fdd{f}{\xi} - f \fdd{\fs}{\xi} +  \gs \fdd{g}{\xi} - g \fdd{\gs}{\xi}\right]^K_{-K}   \sim \ii \laexp\int_{-K}^K (\gs^2- \fs^2 )\, \d \xi.\label{eig0}
\end{multline}
This is the equation which determines the exponentially small correction to the eigenvalue $\laexp$. To find the boundary terms we need to examine the far field more carefully.

\subsection{Boundary condition on a finite domain}
 Consider first $f$.
 We write $f = \fs + \fb$, where  $\fb$ is the correction due to the fact that $\fs$ does not satisfy the boundary conditions.
Then, following the earlier work of~\cite{jon1}, we have
in the far field with $\rho = G \xi$,
\beq
\fs = \Asf \ee^{\ii \phi_2/G}  + \Bsf \ee^{-\ii \phi_2/G} , \qquad\fb = A \ee^{\ii \phi_2/G} + B \ee^{-\ii \phi_2/G},
\label{WKBAreg}
\eeq
where
\beq
\phi_2'= \sqrt{\frac{\rho^2}{4} -1}, \qquad
\ii \la_1  A + 2 \ii \phi_2' A' + \ii \phi_2'' A = \ii G A'', \qquad
\ii  \la_1 B - 2 \ii \phi_2' B' - \ii \phi_2'' B = \ii G B'',\label{Ampeqn}
\eeq
and $\las =  \la_1  G$.  Note  that $\Asf$ and $\Bsf$ are given, but $A$ and $B$ need to be determined.
Expanding
\beq
A = \sum_{n=0}^\infty A_n(\rho) (\ii G)^n, \qquad
B = \sum_{n=0}^\infty B_n(\rho) (-\ii G)^n,\label{WKBAreg2}
\eeq
substituting into Eq.~(\ref{Ampeqn}), and equating coefficients of powers of $G$ gives at leading order
\[
\frac{A_0'}{A_0} = - \frac{(\phi_2'' +\la_1)}{2 \phi_2'}, \qquad
\frac{B_0'}{B_0} = -\frac{(\phi_2''-\la_1)}{2 \phi_2'},
\]
so that
\[
A_0= \frac{\aF}{(\rho^2-4)^{1/4}}\left(\frac{\rho-\sqrt{\rho^2-4}}{\rho+\sqrt{\rho^2-4}}\right)^{\la_1/2} ,\qquad
B_0 =  \frac{\bF}{(\rho^2-4)^{1/4}} \left(\frac{\rho+\sqrt{\rho^2-4}}{\rho-\sqrt{\rho^2-4}}\right)^{\la_1/2},
\]
for some constants $\aF$ and $\bF$.
At the next order
\[  \la_1  A_1 + 2 \phi_2' A_1' +  \phi_2'' A_1 = A_0'', \qquad
-\la_1 B_1 + 2  \phi_2' B_1' +  \phi_2'' B_1 = B_0''.\]
Substituting for $\phi_2$, $A_0$ and $B_0$, and solving gives
\beqas
  A_1
  & = & \frac{\aF}{(\rho^2-4)^{1/4}}\left(\frac{\rho-\sqrt{\rho^2-4}}{\rho+\sqrt{\rho^2-4}}\right)^{\la_1/2} \frac{(24(2\la_1^2-1)\rho + (1-12\la_1^2)\rho^3 - 48 \la_1 \sqrt{\rho^2-4})}{48(\rho^2-4)^{3/2}},\\
 B_1
  & = & \frac{\bF}{(\rho^2-4)^{1/4}}\left(\frac{\rho+\sqrt{\rho^2-4}}{\rho-\sqrt{\rho^2-4}}\right)^{\la_1/2} \frac{(24(2\la_1^2-1)\rho + (1-12\la_1^2)\rho^3 + 48 \la_1 \sqrt{\rho^2-4})}{48(\rho^2-4)^{3/2}},
 \eeqas
 where we fix the constants of integration by requiring that $A \leftrightarrow B$ as we circle the branch point $\rho=2$.
 Continuing in this way, we find that
 \beqa
 A & \sim & \aF \rho^{-1/2-\la_1}(1  + \ii G  \mu_1 - \mu_2G^2+\cdots), \label{Ainf}\\
 B & \sim & \bF  \rho^{-1/2+\la_1}(1  - \ii G  \mu_1 +  \mu_2G^2+\cdots),\label{Binf}
 \eeqa
as $\rho \ra \infty$,  where
 \[ \mu_1 = \frac{(1-12 \la_1^2)}{48}, \qquad \mu_2 = \frac{\la_1(1-4 \la_1^2)}{48}.\]
  A similar asymptotic behaviour must hold for $\Asf$, $\Bsf$, so that
  \beqa
 \Asf & \sim & \asf  \rho^{-1/2-\la_1}(1  + \ii G  \mu_1  - \mu_2G^2+\cdots), \label{Areg}\\
 \Bsf & \sim & \bsf  \rho^{-1/2+\la_1}(1  - \ii G  \mu_1 +  \mu_2G^2+\cdots)\label{Breg}
 \eeqa
as $\rho \ra \infty$. As we approach the turning point $\rho =  2$,
\[ A \sim \frac{\aF}{(4(\rho-2))^{1/4}} ,
\quad
B \sim \frac{\bF}{(4(\rho-2))^{1/4}}.
\]
Matching with the turning point region gives
\[ \aF \ii = \bF,\]
which ensures that the extra contribution due to the reflection back from the boundary is exponentially small in the near field.

The boundary condition gives
\begin{multline*}
\lefteqn{\ii \phi_2' (A+\Asf) \ee^{\ii \phi_2/G} -  \ii \phi_2' (B+\Bsf) \ee^{-\ii \phi_2/G}
+ G(A'+\Asf') \ee^{\ii \phi_2/G} +G(B'+\Bsf') \ee^{-\ii \phi_2/G}} \\
=
\frac{\ii KG}{2} \left((A+\Asf) \ee^{\ii \phi_2/G} + (B+\Bsf) \ee^{-\ii \phi_2/G} \right).
\end{multline*}
Equations (\ref{Ainf})-(\ref{Binf}) show that $A'  = O(A/K)$ for large $K$, so that the term $AK$ dominates $A'$ by a factor of $K^2$. Neglecting the third and fourth terms on the left-hand side gives
\[ \left(\phi_2' - \frac{KG}{2}\right)(A+\Asf) \ee^{\ii \phi_2/G} = \left(\phi_2' + \frac{KG}{2}\right)(B+\Bsf) \ee^{-\ii \phi_2/G},\]
so that
\beqas
\ee^{2 \ii \phi_2/G} \frac{\sqrt{(KG)^2-4} - KG }{\sqrt{(KG)^2-4}+KG }&=&\frac{B+\Bsf}{A+\Asf} .
\eeqas
We now assume that $KG$ is large so that we can use the asymptotic behaviour of Eqs.~(\ref{Ainf})-(\ref{Breg}) to evaluate the right-hand side, giving
\beqas
\frac{\ii  \aF+\bsf}{\aF+\asf} &\sim&-\frac{\ee^{2 \ii \phi_2(KG)/G}}{(KG)^{2+2\la_1}} \left(\frac{1+\ii G \mu_1-\mu_2G^2+\cdots}{1 - \ii G \mu_1+ \mu_2G^2+\cdots}\right)= -S,
\eeqas
say. Then
\beq
\aF  \sim  -\frac{\bsf + \asf S}{\ii + S},\qquad
\bF  \sim   -\ii\frac{\bsf + \asf S}{\ii + S}.\label{aFbF}
\eeq
As $KG \ra \infty$ the behaviour of $S$ (and therefore $\aF$ and $\bF$) crucially depends on whether $\la_1$ is greater or less than $-1$. For $\la_1>-1$, $S \ra 0$ as $KG \ra \infty$ and
\beq
\aF  \sim  \ii \bsf ,\qquad
\bF  \sim   -\bsf.\label{aFbF0}
\eeq
For $\la_1<-1$, $S \ra \infty$ as $KG \ra \infty$ and
\beq
\aF  \sim  -\asf,\qquad
\bF  \sim   -\ii \asf .
\label{aFbFinf}
\eeq
Now, for large $KG$, we can evaluate the boundary terms in Eq.~(\ref{eig0}) as
\begin{eqnarray*}
  \lefteqn{\left.\fs \fdd{f}{\xi} - f \fdd{\fs}{\xi}\right|_{\rho = KG}  =
    \left.\fs \fdd{\fb}{\xi} - \fb \fdd{\fs}{\xi}\right|_{\rho = KG}}&&\\
  &\sim& G \left( \Asf \ee^{\ii \phi_2/G}+\Bsf \ee^{-\ii \phi_2/G}\right) \left( \frac{\ii  \phi_2'}{G} \left(A\ee^{\ii \phi_2/G} - B\ee^{-\ii \phi_2/G}\right) +A'\ee^{\ii \phi_2/G} + B'\ee^{-\ii \phi_2/G} \right)
\\ && \mbox{ } - G \left(A\ee^{\ii \phi_2/G} + B\ee^{-\ii \phi_2/G}\right)
\left( \frac{\ii  \phi_2'}{G} \left(\Asf\ee^{\ii \phi_2/G} - \Bsf\ee^{-\ii \phi_2/G}\right) +\Asf'\ee^{\ii \phi_2/G} + \Bsf'\ee^{-\ii \phi_2/G} \right)\\
&\sim&  -2\ii \phi_2'( \Asf  B -\Bsf A)\\
 &\sim&-\ii(\asf (1+\ii \mu_1 G +\cdots)\bF(1-\ii \mu_1 G+\cdots)-\bsf(1-\ii \mu G+\cdots) \aF(1+\ii \mu G+\cdots))\\
 &=& -\ii(\asf \bF-\bsf \aF)\left(1 -(\ii \mu_1 G -  \mu_2 G^2+\cdots)^2\right).
\end{eqnarray*}
A similar calculation on $g$  shows that, when $\la_1$ is real, 
\beqas
\left.\gs \fdd{g}{\xi} - g \fdd{\gs}{\xi} \right|_{\rho = KG} & \sim&  - \ii (\asg \bG-\bsg \aG)\left(1 - (-\ii \mu_1 G - \mu_2 G^2+\cdots)^2\right),
\eeqas
where
\[
\aG  = \ii\frac{  \asg +   \bsg S^*}{-\ii   + S^*}
,\qquad
\bG  = -\frac{  \asg +   \bsg S^*}{-\ii   + S^*},
\]
so that
\begin{align}
  \aG &\sim -  \asg,& \bG &\sim - \ii \asg, & \la_1&>-1 ,\\
  \aG &\sim \ii  \bsg,&  \bG &\sim - \bsg, & \la_1&<-1.
\end{align}
A similar calculation of the boundary layer at $-K$ gives, finally,
\begin{multline}
  \left[ \fs \fdd{f}{\xi} - f \fdd{\fs}{\xi} +  \gs \fdd{g}{\xi} - g \fdd{\gs}{\xi}\right]^K_{-K} \sim \\
  \begin{cases}
    \begin{split}
      &2\ii \bsf(\asf +  \ii \bsf)\left(1 -(\ii \mu_1 G -  \mu_2 G^2+\cdots)^2\right)\\
&\mbox{ }- 2\ii\asg (-  \ii \asg+\bsg  )\left(1 - (-\ii \mu_1 G - \mu_2 G^2+\cdots)^2\right)
\end{split} & \mbox{ if } \la_1>-1,\\[5mm]
\begin{split}&  -2\ii\asf(-  \ii \asf+\bsf )\left(1 -(\ii \mu_1 G -  \mu_2 G^2+\cdots)^2\right)\\
&\mbox{ }  + 2 \ii \bsg (\asg +  \ii  \bsg)\left(1 - (-\ii \mu_1 G - \mu_2 G^2+\cdots)^2\right)
\end{split} & \mbox{ if } \la_1<-1.
\end{cases}
\end{multline}
We will see that $\asg = \left(\bsf\right)^*$ and $\bsg = \left(\asf\right)^*$ so that the right-hand side is real.

\subsection{Eigenvalues}
We now apply the general methodology to each of the eigenvalues in turn.
Since the approximate eigenfunctions $\fs$ and $\gs$ are given in terms of the steady state solution $\Vs$, to identify the coefficients $\asf$, $\asg$, $\bsf$ and $\bsg$ that appear in the boundary terms it is useful to recall the behaviour of $\Vs$ in the far field, which was determined in \cite{jon1}. There we found that 
\beqa
\Vs & \sim &\al  \ee^{\ii \phi_2/G} \sum_{n=0}^\infty A_n(\rho) (\ii G)^n+
\beta\ee^{-\ii \phi_2/G}  \sum_{n=0}^\infty A_n(\rho) (-\ii G)^n,\qquad
A_0(\rho)  =  \frac{2^{1/2}a_0}{(\rho^2-4)^{1/2}}, \label{Vss}
\eeqa
where
\beqas
a_0 & = & 12^{1/4},\\
\al &=& \frac{ \ee^{\ii \pi/4} \ee^{-\pi/2G}}{1 - \ii \nu \ee^{2 \ii \phi_2(KG)/G}} = \ee^{\ii \pi/4} \ee^{-\pi/2G}+\frac{\ii \nu\ee^{2 \ii \phi_2(KG)/G} 
  \ee^{\ii \pi/4} \ee^{-\pi/2G}}{1 - \ii \nu \ee^{2 \ii \phi_2(KG)/G}},\\   \beta &=& -\nu \alpha \ee^{2 \ii \phi_2(KG)/G} = -\frac{\nu \ee^{2 \ii \phi_2(KG)/G}
 \ee^{\ii \pi/4} \ee^{-\pi/2G}}{1 - \ii \nu \ee^{2 \ii \phi_2(KG)/G}}, \\
\nu &\sim& \frac{KG-\sqrt{(KG)^2-4} }{KG+\sqrt{(KG)^2-4} } \sim \frac{1}{(KG)^2}.
\eeqas

\subsubsection{The eigenvalue $\las = 2G$}
\label{la2}
\begin{figure}
  \begin{center}
\begin{overpic}[width=0.95\textwidth]{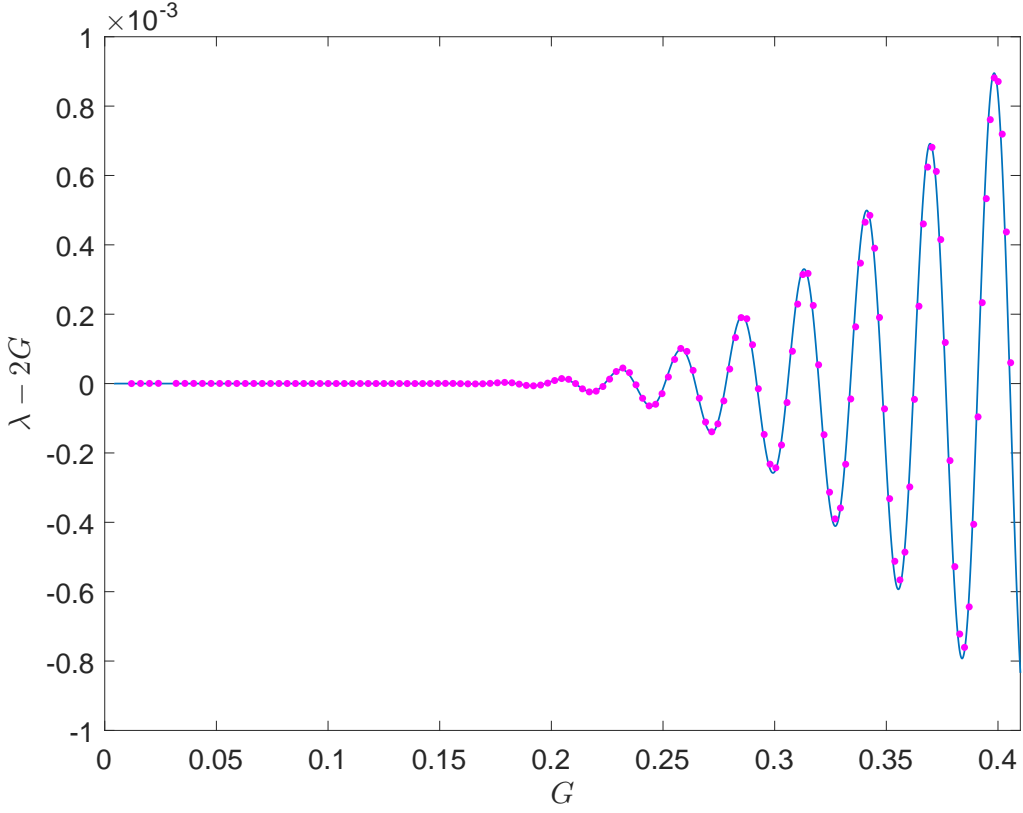}
\end{overpic}
    \end{center}
  \caption{Comparison between numerical and the asymptotic predictions, for $K=20$. The solid blue curve corresponds to Eq.~(\ref{asy2G}), while
  the purple dots to the numerical solution. }
  \label{figla2G}
\end{figure}

In terms of $f$ and $g$ the approximate eigenfunctions are
\beq
\fs  =  \ii \Vs + G \left(\frac{\Vs}{\sigma} + \xi \fdd{\Vs}{\xi} - \frac{\ii G \xi^2 \Vs}{2}\right),\quad
\gs  =  -\ii \Vs^*+ G \left(\frac{\Vs^*}{\sigma} + \xi \fdd{\Vs^*}{\xi} + \frac{\ii G \xi^2 \Vs^*}{2}\right).\label{FG2G}
\eeq
These satisfy the equations exactly  so that
$\Rf=\Rg=0$. The perturbation to the eigenvalue arises solely because of the finiteness of the domain, since $\fs$ and $\gs$ do not satisfy the boundary conditions.
From the known expansion, Eq.~(\ref{Vss}) of the steady state solution, we need to identify the amplitude coefficients $\asf$, $\bsf$ in the WKB expansion [cf. Eqs.~(\ref{WKBAreg})-(\ref{WKBAreg2})]. The easiest way to do this is to compare the two representations of $\fs$ and $\gs$ as $\rho \ra \infty$.
Comparing  Eqs.~(\ref{Vss})-(\ref{FG2G}) with  Eqs.~(\ref{Areg})-(\ref{Breg}) as $\rho \ra \infty$ gives
\beqas
\asf\left(1-\ii \frac{47 G}{48}\right) 
&\sim& - \ii \sqrt{2}a_0 \alpha \left( 1 - \ii \frac{95 G}{48} + \cdots\right) ,
\\
\bsf\left(1+\ii \frac{47 G}{48}\right)  &\sim& - \ii \sqrt{2}a_0 \beta \left( 1 -\ii \frac{G}{48} + \cdots\right) ,
\eeqas
since
\[ \mu_1 =  \frac{(1-12 \la_1^2)}{48} = -\frac{47}{48}.\]
Thus, 
\[ \asf \sim -\ii\sqrt{2}a_0   \al\left( 1- \ii G\right)
, \qquad
\bsf \sim -\ii \sqrt{2}a_0  \beta \left(1- \ii G \right), \qquad
\asg \sim \left(\bsf \right)^*, \qquad
\bsg \sim \left(\asf \right)^* .\]
Then
\beqas
\left[ \fs \fdd{f}{\xi} - f \fdd{\fs}{\xi} +  \gs \fdd{g}{\xi} - g \fdd{\gs}{\xi}\right]^K_{-K} 
&\sim&  2\ii \bsf(\asf +  \ii \bsf)- 2\ii\asg (-  \ii \asg+\bsg  ) \\
   & \sim & -8 a_0^2 \nu \ee^{-\pi/G}
\re\left(( 1- \ii G)^2\ee^{2 \ii \phi_2(KG)/G}  \right),
\eeqas
as $\nu \ra 0$.
Evaluating the right-hand side of Eq.~(\ref{eig0}) gives
\beqas
 \ii \laexp \int_{-\infty}^\infty 
   (\gs^2- \fs^2 )\, \d \xi
   & = &  \ii \laexp \int_{-\infty}^\infty (\gs-\fs)(\gs+\fs)
\, \d \xi\\
   & = & -2 G   \laexp \int_{-\infty}^\infty \left(- 2 \Vs+ G^2 \xi^2 \Vs\right)  \left(\frac{\Vs}{\sigma} + \xi \fdd{\Vs}{\xi}  \right)
\, \d \xi\\
& = &- 2 G   \laexp \int_{-\infty}^\infty - 2\Vs^2  \left(\frac{1}{\sigma}- \frac{1}{2}  \right) +  G^2 \xi^2 \Vs^2\left(\frac{1}{\sigma} - \frac{3}{2}\right)
\, \d \xi,
\eeqas
since
\[
\int_{-\infty}^\infty \xi \fdd{\Vs}{\xi} \Vs\, \d \xi =
- \frac{1}{2}\int_{-\infty}^\infty  \Vs^2\, \d \xi,\qquad
\int_{-\infty}^\infty \xi^3 \fdd{\Vs}{\xi} \Vs\, \d \xi =
- \frac{3}{2}\int_{-\infty}^\infty \xi^2 \Vs^2\, \d \xi.
\]
The dominant contribution to these integrals is from the near field \cite{jon1}. Using the asymptotic expansion of $\Vs$ in powers of $G$ \cite{jon1} gives
\beq \int_{-\infty}^\infty \Vs^2\, \d \xi =\frac{\sqrt{3}\pi}{2}+  \frac{\sqrt{3}\, \pi^3G^2}{128} + O(G^4), \qquad
\int_{-\infty}^\infty \xi^2\Vs^2\, \d \xi =\frac{\sqrt{3}\pi^3}{32}+O(G^2),\label{intV2}
\eeq
so that Eq.~(\ref{eig0}) becomes
\beqas
- 8 a_0^2 \nu \ee^{-\pi/G}
\re\left(( 1- \ii G)^2\ee^{2 \ii \phi_2(KG)/G}  \right)
& = & - 2 \sqrt{3}G \pi  \laexp\left( -   \left(\frac{1}{\sigma}- \frac{1}{2}  \right) +  G^2 \frac{\pi^2}{32}\left(\frac{1}{\sigma} - \frac{3}{2}\right)\right)\\
& \sim &    \frac{\sqrt{3}G^3 \pi^3  \laexp}{16},
\eeqas
since $\sigma$ is exponentially close to 2.
Thus, the correction to the eigenvalue is
\beqa
\laexp 
& \sim &  -\frac{256  \nu \ee^{-\pi/G}}{  G^3 \pi^3 }
\re\left(( 1- \ii G)^2\ee^{2 \ii \phi_2(KG)/G}  \right).\label{asy2G}
\eeqa
A comparison between Eq.~(\ref{asy2G}) and the numerically calculated eigenvalue  for $K=20$ is shown in Fig.~\ref{figla2G}. This shows that our oscillatory
correction excellently captures the correction due to the finiteness
of the domain around the dominant $\las = 2 G$.

\subsubsection{The eigenvalue $\las = G$}
In terms of $f$ and $g$, the approximate  eigenfunctions are
\beq
\fs  =  \fdd{\Vs}{\xi} - \frac{\ii G \xi \Vs}{2},\qquad
\gs  =  \fdd{\Vs^*}{\xi} + \frac{\ii G \xi \Vs^*}{2}.\label{FGG}
\eeq
Again these satisfy the equations exactly, so that
$\Rf=\Rg=0$, and the perturbation to the eigenvalue arises solely because of the finiteness of the domain.

\begin{figure}
  \begin{center}
\begin{overpic}[width=0.95\textwidth]{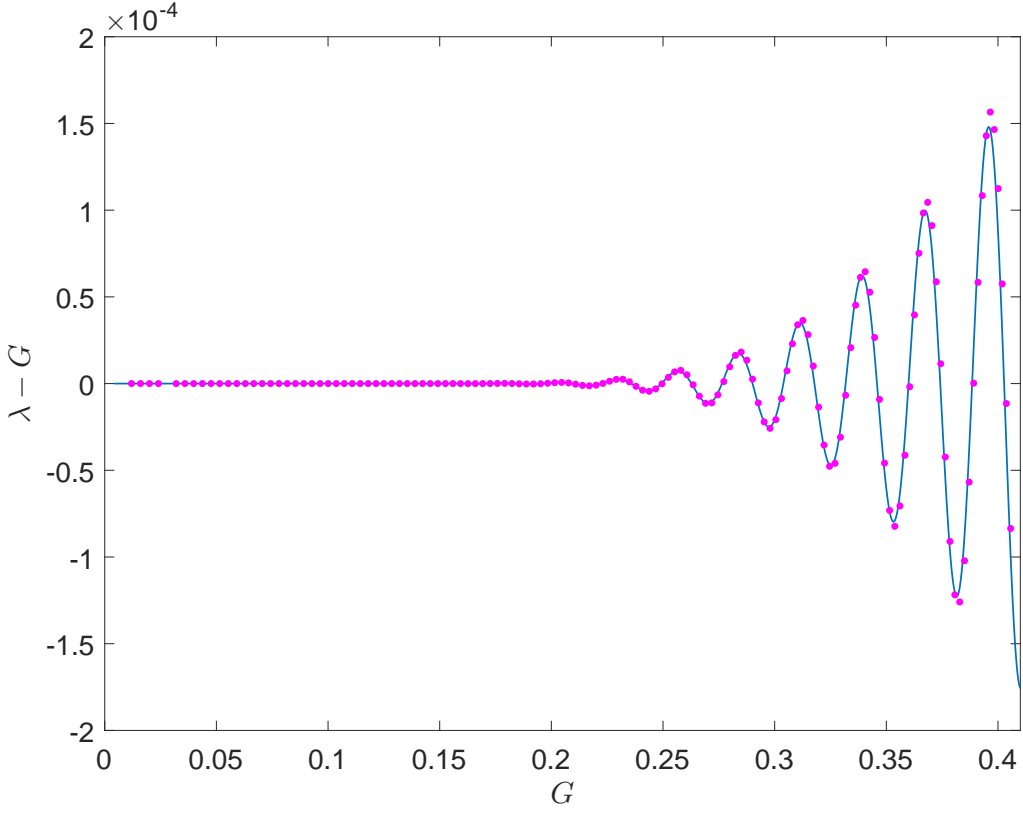}
\end{overpic}
    \end{center}
  \caption{Comparison between numerical and the asymptotic predictions, for $K=20$. The solid blue curve corresponds to Eq.~(\ref{asyG}), while 
  the purple dots pertain to
    numerical solution.}
  \label{figlaG}
\end{figure}

Comparing Eqs.~(\ref{FGG}) with  Eqs.~(\ref{Areg})-(\ref{Breg}) as $\rho \ra \infty$ gives
\[ \asf\left(1-\ii \frac{11 G}{48}\right) 
\sim - \ii \sqrt{2}a_0  \alpha \left( 1 - \ii \frac{23 G}{48} + \cdots\right) ,
\quad
\bsf\left(1+\ii \frac{11 G}{48}\right)  \sim - \ii  \sqrt{2}a_0 \beta \left( 1 -\ii \frac{G}{48} + \cdots\right) ,
\]
since
\[ \mu_1 =  \frac{(1-12 \la_1^2)}{48} = -\frac{11}{48}.\]
Thus
\[ \asf \sim -\ii \sqrt{2}a_0  \al\left( 1-\frac{ \ii  G}{4}\right)
, \quad
\bsf \sim -\ii  \sqrt{2}a_0 \beta \left(1- \frac{ \ii  G}{4} \right), \quad
\asg \sim \bsf^*, \quad
\bsg \sim  \asf^*.\]
Then,
\beqas
\left[ \fs \fdd{f}{\xi} - f \fdd{\fs}{\xi} +  \gs \fdd{g}{\xi} - g \fdd{\gs}{\xi}\right]^K_{-K}
   & \sim & -8 a_0^2 \nu \ee^{-\pi/G}
\re\left(( 1- \ii G/4)^2\ee^{2 \ii \phi_2(KG)/G}  \right).
\eeqas
Evaluating the right-hand side of Eq.~(\ref{eig0}) gives
\beqas
 \ii \laexp \int_{-\infty}^\infty 
   (\gs^2- \fs^2 )\, \d \xi
      & = & -2 G  \laexp \int_{-\infty}^\infty  \xi \Vs \fdd{\Vs}{\xi}
\, \d \xi
 =  G  \laexp \int_{-\infty}^\infty   \Vs^2
\, \d \xi.
  \eeqas
Using Eq.~(\ref{intV2}) gives
\beqas
- 8 a_0^2 \nu \ee^{-\pi/G}
\re\left(( 1- \ii G/4)^2\ee^{2\ii \phi_2(KG)/G} \right)& = &  G   \laexp\frac{\sqrt{3}\pi}{2},
\eeqas
i.e.,
\beqa
\laexp
&\sim& -\frac{32 \nu \ee^{-\pi/G}}{G\pi}
\re\left(( 1- \ii G/4)^2\ee^{2\ii \phi_2(KG)/G} \right)\label{asyG}
\eeqa
In this case, a comparison between Eq.~(\ref{asyG}) and the numerically calculated eigenvalue  for $K=20$ is shown in Fig.~\ref{figlaG}. Once again,
very good agreement is observed with the numerical finite-domain-induced
oscillations, even for values of $G$ that are quite high (i.e., near
$0.5$).

\subsubsection{The eigenvalue $\las = -2G$}
\label{sec:laM2G}
In terms of $f$ and $g$ the approximate eigenfunctions are
\beq
\fs  =  \ii \Vs - G \left(\frac{\Vs}{\sigma} + \xi \fdd{\Vs}{\xi} + \frac{\ii G \xi^2 \Vs}{2}\right),\quad
\gs  =  -\ii \Vs^*- G \left(\frac{\Vs^*}{\sigma} + \xi \fdd{\Vs^*}{\xi} - \frac{\ii G \xi^2 \Vs^*}{2}\right).\label{FGm2G}
\eeq
This time, the approximate eigenfunctions do not satisfy the equation exactly, but with an exponentially small residual.
We find
\[
\Rf  =  - 4 \ii G^2 \Vs\left(\frac{1}{2} - \frac{1}{\sigma}\right),\qquad
\Rg  =    4 \ii G^2 \Vs^*\left(\frac{1}{2} - \frac{1}{\sigma}\right),
\]
so that
\beqas
\int_{-\infty}^\infty \Rf \fs + \Rg \gs\, \d \xi & = &  4 G^2 \left(\frac{1}{2} - \frac{1}{\sigma}\right)\int_{-\infty}^\infty\left(2\Vs^2 - G^2 \xi^2 \Vs^2\right) \, \d \xi\\
& = & 4 G^2 \left(\frac{1}{2} - \frac{1}{\sigma}\right)\left( \sqrt{3} \pi  +\frac{\sqrt{3}\pi^3 G^2}{64}
- \frac{\sqrt{3}\pi^3 G^2}{32} + \cdots\right) \\
& = & 4 G^2 \left(\frac{1}{2} - \frac{1}{\sigma}\right)\left( \sqrt{3} \pi
- \frac{\sqrt{3}\pi^3 G^2}{64} + \cdots\right).
\eeqas
Unfortunately, for $\las = -2 G$ we will find that we will need to know more than the leading-order behaviour of $\asf$ and $\bsf$ in order to find the leading-order approximation to $\laexp$.
Comparing Eqs.~(\ref{FGm2G}) with Eqs.~(\ref{Areg})-(\ref{Breg}) at infinity, including higher-order terms in both expansions (see \cite{jon1}), gives 
\beqas
\asf\left(1+\ii \mu_1 G - \mu_2 G^2- \ii \mu_3G^3\right) 
&\sim& - \ii \sqrt{2}a_0 \kappa \alpha \left( 1 +  \frac{\ii G}{48} + \frac{2021 \ii G^3}{1658880}+ \cdots\right) ,\\
\bsf\left(1-\ii \mu_1 G+\mu_2 G^2+\ii \mu_3G^3\right)  &\sim& - \ii  \beta \sqrt{2}a_0 \kappa\left( 1 +\ii \frac{95G}{48}  - \frac{17 G^2}{24} +
\frac{23899\ii G^3}{1658880} +\cdots\right),
\eeqas
where
\[ \kappa \sim 1 - \left(\frac{1+12\pi^2}{4608}\right)G^2 + 0.0152 G^4 + \cdots.
\]
Since
\[ \mu_1 = \frac{(1-12 \la_1^2)}{48} = -\frac{47}{48},
\quad \mu_2 = \frac{\la_1(1-4 \la_1^2)}{48}=\frac{5}{8},
\quad \mu_3 =-\frac{450581}{1658880},
\]
we find 
\beqas
\asf &\sim& -\ii  \al\sqrt{2}a_0 \kappa \left( 1+ \ii G - \frac{17 G^2}{48} + \frac{\ii G^3}{128}+\cdots\right)
, \\
\bsf &\sim& -\ii  \beta \sqrt{2}a_0 \kappa \left( 1+ \ii G - \frac{17 G^2}{48} + \frac{\ii G^3}{128}+\cdots\right)
\eeqas
as well as
\[
\asg = \left(\bsf \right)^*,\qquad  \bsg = \left(\asf \right)^*.\]
Then, as $KG \ra \infty$,
\beqas
\lefteqn{ \left[ \fs \fdd{f}{\xi} - f \fdd{\fs}{\xi} +  \gs \fdd{g}{\xi} - g \fdd{\gs}{\xi}\right]^K_{-K}} \hspace{1cm}&& \\
 &\sim& -2\ii\asf(-  \ii \asf )\left(1 -(\ii \mu_1 G -  \mu_2 G^2+\cdots)^2\right)+ 2 \ii \bsg (  \ii  \bsg)\left(1 - (-\ii \mu_1 G - \mu_2 G^2+\cdots)^2\right)\\
&\sim& 8 \sqrt{3}\ii \ee^{-\pi/G}   \kappa^2\left( 1+ \ii G - \frac{17 G^2}{48} + \frac{\ii G^3}{128}+\cdots\right)^2 \left(1 -(\ii \mu_1 G -  \mu_2 G^2+\cdots)^2\right)\\
&& \mbox{ }- 8 \sqrt{3}\ii \ee^{-\pi/G}   \kappa^2\left( 1- \ii G - \frac{17 G^2}{48} - \frac{\ii G^3}{128}+\cdots\right)^2 \left(1 -(-\ii \mu_1 G -  \mu_2 G^2+\cdots)^2\right)\\
&  \sim &   - 32\sqrt{3}\,  \ee^{-\pi/G} \kappa^2 \left(G
+ \frac{G^3}{2304} + \cdots\right).
\eeqas
 Evaluating the right-hand side of Eq.~(\ref{eig0}) gives
\beqas
 \ii \laexp \int_{-\infty}^\infty 
   (\gs^2- \fs^2 )\, \d \xi
    & = & 2 G   \laexp \int_{-\infty}^\infty \left(- 2 \Vs+ G^2 \xi^2 \Vs\right)  \left(\frac{\Vs}{\sigma} + \xi \fdd{\Vs}{\xi}  \right)
\, \d \xi\\
& = & 2 G   \laexp \int_{-\infty}^\infty - 2\Vs^2  \left(\frac{1}{\sigma}- \frac{1}{2}  \right) +  G^2 \xi^2 \Vs^2\left(\frac{1}{\sigma} - \frac{3}{2}\right)
\, \d \xi,
\eeqas
after integrating by parts.
Using  (\ref{intV2}) we find   that (\ref{eig0}) becomes
\beqa
\lefteqn{- 32\sqrt{3}\,  \ee^{-\pi/G} \kappa^2 G \left(1+\frac{G^2}{2304}  + \cdots\right)
 \mbox{}+4 G^2 \left(\frac{1}{2} - \frac{1}{\sigma}\right)\left( \sqrt{3} \pi  - \frac{\sqrt{3}\pi^3 G^2}{64}\right)}\hspace{5cm}&&\non \\
&=&   2\sqrt{3}G \pi  \laexp\left( -   \left(\frac{1}{\sigma}- \frac{1}{2}  \right) +  G^2 \frac{\pi^2}{32}\left(\frac{1}{\sigma} - \frac{3}{2}\right)\right)\non\\
& \sim &  -\frac{ 2\sqrt{3}G^3 \pi^3 \laexp}{32} \label{laexpm2Geqn}
\eeqa
Now,
since (for $\nu \sim 0$) the relation between $\sigma$ and $G$ is (see \cite{jon1}) 
\[G\left(\frac{1}{2} - \frac{1}{\sigma}\right) \left(\frac{\sqrt{3} \pi}{4} + \frac{\sqrt{3} \pi^3G^2}{256} + \cdots \right) = 2 \sqrt{3} \kappa^2 \ee^{-\pi/G},\]
we find the leading terms on the left-hand side of (\ref{laexpm2Geqn}) vanish. This is the reason we needed to include the higher-order corrections; these  give the correction to the eigenvalue as 
\beq
\laexp\sim  G\,  \left(1- \frac{2}{\sigma}\right)
 \left( 1 + \frac{1}{72 \pi^2}\right)+\cdots.\label{asyM2G}
\eeq

\begin{figure}
 \begin{center}
\begin{overpic}[width=0.95\textwidth]{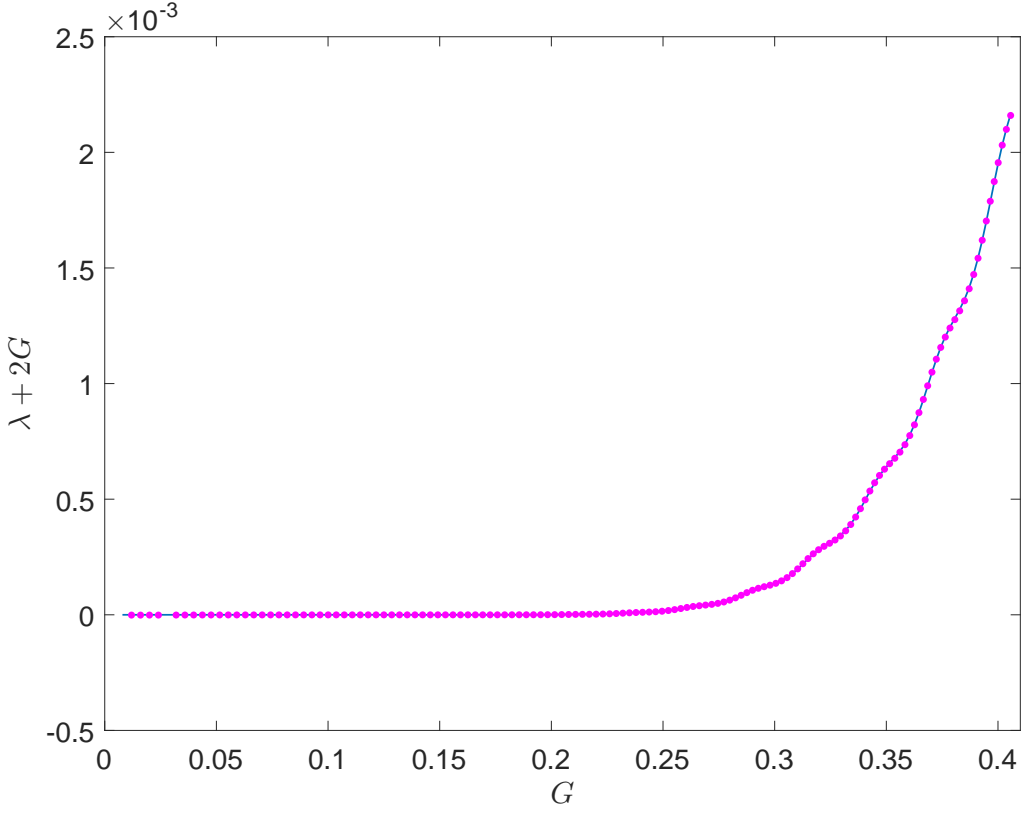}
\end{overpic}\vspace{6mm}
    \end{center}
  \caption{Asymptotic prediction (\ref{asyM2G}) (blue) compared to numerical solution (purple) for the case of the eigenvalue with $\las = -2 G$. }
 \label{figM2Gasy}
\end{figure}

A comparison between (\ref{asyM2G}) and the numerically calculated eigenvalue  for $K=20$ is shown in Fig.~\ref{figM2Gasy}.
Note that although we derived (\ref{asyM2G}) in the limit $K \ra \infty$, $\nu \ra 0$, when we plot it in  Fig.~\ref{figM2Gasy} we use the finite-domain approximation to $\sigma$ as a function of $G$.
We can see
that for this eigenvalue we do not purely observe the oscillatory effect induced
by the finite nature of the domain as in the two previous cases.
Rather, the relevant correction incorporates 
also the deviation from the exact scaling symmetry (and hence
from the symmetry of the eigenvalue pair at $\pm 2 G$) which provides
the monotonic portion of the relevant correction.

\subsubsection{The eigenvalue $\las = -G$}
In terms of $f$ and $g$, the approximate eigenfunctions are
\beq
\fs  =  \fdd{\Vs}{\xi} + \frac{\ii G \xi \Vs}{2},\qquad
\gs  =  \fdd{\Vs^*}{\xi} - \frac{\ii G \xi \Vs^*}{2}.
\label{evG}
\eeq
These satisfy the equations exactly, so that $\Rf=\Rg=0$.
However, they do not satisfy the correct radiation condition at infinity. In the finite domain context, the perturbation of the eigenvalue arises from the boundary terms in Eq.~(\ref{eig0}).

Comparing Eq.~(\ref{evG}) with
 Eqs.~(\ref{Areg})-(\ref{Breg}) as $\rho \ra \infty$ gives
\[ \asf\left(1-\ii \frac{11 G}{48}\right) 
\sim  \ii \sqrt{2}a_0  \alpha \left( 1 + \ii \frac{ G}{48} + \cdots\right) ,
\quad
\bsf\left(1+\ii \frac{11 G}{48}\right)  \sim  \ii  \sqrt{2}a_0 \beta \left( 1+\ii \frac{23 G}{48} + \cdots\right),
\]
since
\[ \mu_1 =  \frac{(1-12 \la_1^2)}{48} = -\frac{11}{48}.\]
Thus,
\[ \asf \sim \ii \sqrt{2}a_0  \al\left( 1+\frac{ \ii  G}{4}\right)
, \quad
\bsf \sim \ii  \sqrt{2}a_0 \beta \left(1+ \frac{ \ii  G}{4} \right), \quad
\asg \sim \left(\bsf \right)^*, \quad
\bsg \sim  \left(\asf \right)^*.\]
When $\la_1 = -1$, $S\ra  \ee^{2\ii \phi_2/G}$ as $\nu \ra 0$, so we need to use the full expressions 
\beq
\aF  \sim  -\frac{\bsf + \asf S}{\ii + S},\qquad
\bF  \sim   -\ii\frac{\bsf + \asf S}{\ii + S},
\eeq
for $\aF$ and $\bF$.
Then
\beqas
\left[ \fs \fdd{f}{\xi} - f \fdd{\fs}{\xi} +  \gs \fdd{g}{\xi} - g \fdd{\gs}{\xi}\right]^K_{-K} &\sim
& \re\left(-4\ii(\asf \bF-\bsf \aF)\left(1 +  \mu_1^2 G^2+\cdots\right)\right)\\
& \sim &  \re\left(-\frac{8\ii a_0^2(1+\ii G/4)^2 }{\ii + S} 
(\ii \alpha - \beta)(S \alpha +\beta ) \left(1 +  \mu_1^2 G^2+\cdots\right)\right)\\
& \sim &  \re\left(\frac{8\ii a_0^2(1+\ii G/4)^2 }{\ii + S} 
S \ee^{-\pi/G}  \left(1 +  \mu_1^2 G^2+\cdots\right)\right)\\
& \sim &  \re\left(\frac{16\ii  \sqrt{3}(1+\ii G/4)^2 }{1+\ii  \ee^{-2\ii \phi_2/G}} 
 \ee^{-\pi/G} \right)
\eeqas

\begin{figure}
 \begin{center}
\begin{overpic}[width=0.8\textwidth]{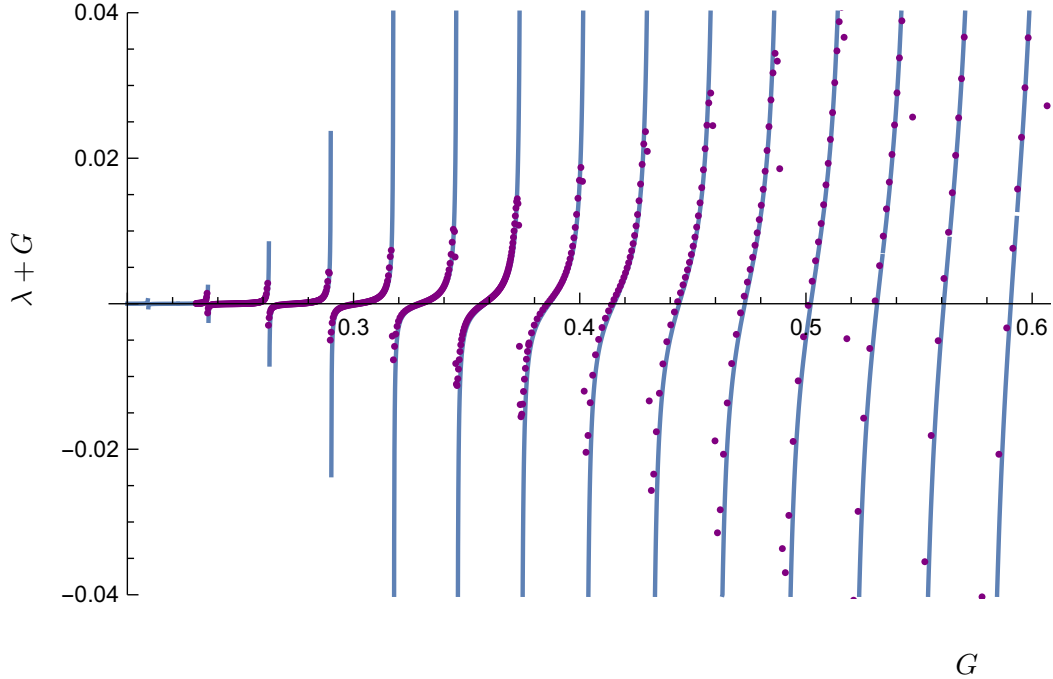}
\put(-5,30){\rotatebox{90}{$\lambda+G$}}
 \put(90,-7){$G$}
\end{overpic}\vspace{6mm}
    \end{center}
  \caption{Asymptotic prediction [cf. Eq.~(\ref{asyMG})] (blue) compared to numerical solution (purple), for the eigenvalues with $\las=-G$ and for $K=20$. }
 \label{figMGasy}
\end{figure}

Since the integrals in Eq.~(\ref{eig}) are dominated by the near field, where $\Vs$ is real,  using the near-field solution in  Eq.~(\ref{eig}) gives
\beqas
 \ii \laexp \int_{-\infty}^\infty 
   (\gs^2- \fs^2 )\, \d \xi
   & = &  \ii \laexp \int_{-\infty}^\infty (\gs-\fs)(\gs+\fs)
\, \d \xi\\
     & = & 2 G  \laexp \int_{-\infty}^\infty  \xi \Vs \fdd{\Vs}{\xi}
\, \d \xi\\
& = &  -G  \laexp \int_{-\infty}^\infty   \Vs^2
\, \d \xi \sim  -G  \laexp\frac{\sqrt{3}\pi}{2}.
  \eeqas
  Thus the correction to the eigenvalue is
  \beq \laexp \sim -\re\left(\frac{32\ii (1+\ii G/4)^2 }{\pi G(1+\ii  \ee^{-2\ii \phi_2/G})}  
  \ee^{-\pi/G} \right).\label{asyMG}
  \eeq
A comparison between Eq.~(\ref{asyMG}) and the numerically calculated eigenvalue  for $K=20$ is shown in Fig.~\ref{figMGasy}. A key feature to observe here is
the presence of vertical asymptotes in this exponentially small (in $1/G$)
correction. These represent the reason for the jumps observed in 
Fig.~\ref{figMGasy}. Indeed, it is relevant to note that a
particularly careful observation of the orange line in
Fig.~\ref{fig2} will reveal the outcome of these jumps to the
particularly astute reader, as can be discerned, e.g., near
the outermost disconnect of the relevant numerical line.
Despite the fact that our theoretical approximation can no longer
be considered accurate when $\laexp$ becomes large, we can still see
that it very accurately captures our numerical results of 
Fig.~\ref{figMGasy}.

We see from the numerical results that there is a very thin transition region in the vicinity of each asymptote in which the eigenvalue perturbation switches from large and  positive to large and negative. We do not attempt to capture this transition region, which requires a detailed calculation in the vicinity of $S = -\ii$.

\subsubsection{The eigenvalues near $\la=0$}
\label{sec:zeroev}
In terms of $f$ and $g$ the approximate eigenfunctions are
\beq
\fs  =  \ii \Vs, \qquad
\gs  =  -\ii \Vs^*.
\eeq
In fact, these satisfy the equations and boundary conditions exactly,
so that $\la=0$ is an exact eigenvalue even for a finite
domain. However, as we have seen numerically, there is a second
eigenvalue which is exponentially close to zero, which we now
approximate. This analysis does not fit into the general framework of
Section \ref{theory}, but follows a similar methodology, which we now
outline.

We write  
 \beqa
 f & = & \ii \Vs + {\laexp}_1  f_1 + \fexp ,\label{j3}\\
 g & = & -\ii \Vs^* + {\laexp}_1 g_1 + \gexp ,\\
\la & = & {\laexp}_1 + {\laexp}_2,\label{j5}
\eeqa
where ${\laexp}_2 \ll {\laexp}_1$ and
\beqa
\sdd{ f_1}{\xi} +
\sigma |\Vs|^{2\sigma-2}(\Vs^*)^{2} g_1
+ ( \sigma+1)|\Vs|^{2 \sigma}  f_1 
- f_1- \frac{\ii (\sigma-2) G}{2 \sigma} f_1
+ \frac{G^2 \xi^2 }{4} f_1 &=&   \Vs^*,\label{j1}\\
  \sdd{g_1}{\xi} 
+  \sigma |\Vs|^{2\sigma-2}(\Vs)^{2} f _1
+ ( \sigma+1)|\Vs|^{2 \sigma} g_1
-  g_1+ \frac{\ii (\sigma-2) G}{2 \sigma}  g_1 
+ \frac{G^2 \xi^2 }{4} g_1  &=&   \Vs,\label{j2}
\eeqa
with
\[\fdd{f_1}{\xi} = \frac{\ii G \xi f_1}{2}, \qquad \fdd{g_1}{\xi} = - \frac{ \ii G \xi g_1}{2} \qquad \mbox{ at } \xi = K.
\]
Note that the linear operator here is slightly different from (but exponentially close to) that of Eqs.~(\ref{geqn})-(\ref{feqn}), and is chosen so that the 
solvability condition is exactly satisfied, so that we can be sure that
$f_1$, $g_1$ exist: multiplying Eq.~(\ref{j1}) by $\Vs$ and Eq.~(\ref{j2}) by $-\Vs^*$ adding and integrating gives
\beqas
&&\hspace{-2cm}\int_{-K}^{K}\left(\sdd{ f_1}{\xi} +
\sigma |\Vs|^{2\sigma-2}(\Vs^*)^{2} g_1  
+ ( \sigma+1)|\Vs|^{2 \sigma}  f_1
- f_1
- \frac{\ii (\sigma-2) G}{2 \sigma} f_1
+ \frac{G^2 \xi^2 }{4} f_1 \right) \Vs\, \d \xi \\
&& \hspace{-2cm}-
\int_{-K}^{K}\left(
 \sdd{g_1}{\xi} 
+  \sigma |\Vs|^{2\sigma-2}\Vs^{2} f_1
+ ( \sigma+1)|\Vs|^{2 \sigma} g_1 
-  g_1 
+ \frac{\ii (\sigma-2) G}{2 \sigma}  g_1 
+ \frac{G^2 \xi^2 }{4} g_1\right) \Vs^*\, \d \xi\\
& = & \int_{-K}^{K}\left(\sdd{ \Vs}{\xi} 
+|\Vs|^{2\sigma}  \Vs 
- \Vs
- \frac{\ii(\sigma-2)G}{2 \sigma} \Vs
+ \frac{G^2 \xi^2 }{4} \Vs\right) f_1\, \d \xi\\
&&- \int_{-K}^{K}\left(\sdd{\Vs^*}{\xi} 
  + |\Vs|^{2\sigma}\Vs^*
-  \Vs^* + \frac{\ii(\sigma-2)G}{2 \sigma} \Vs^*
+ \frac{G^2 \xi^2 }{4}\Vs^*\right) g_1\, \d \xi
\\
&& + \left[ \Vs \fdd{f_1}{\xi} - f_1 \fdd{\Vs}{\xi} -  \Vs^* \fdd{g_1}{\xi} + g_1 \fdd{\Vs^*}{\xi}\right]^K_{-K}\\
& = &  \left[ \Vs \frac{\ii G \xi f_1}{2} - f_1 \frac{\ii G \xi \Vs}{2} -  \Vs^* \frac{(-\ii G \xi g_1)}{2} + g_1 \frac{(-\ii G \xi \Vs^*)}{2}\right]^K_{-K}
 =  0.
\eeqas
Now, substituting Eqs.~(\ref{j3})-(\ref{j5}) into Eqs.~(\ref{geqn})-(\ref{feqn})  gives
\beqas
\lefteqn{\sdd{ \fexp}{\xi} +
\sigma |\Vs|^{2\sigma-2}\Vs^{2} \gexp
+ ( \sigma+1)|\Vs|^{2 \sigma}   \fexp 
- \fexp- \frac{\ii (\sigma-2) G}{2 \sigma} \fexp
+ \frac{G^2 \xi^2 }{4}  \fexp }\hspace{2cm} && \\
&=&  {\laexp}_1\sigma  |\Vs|^{2\sigma-2}((\Vs^*)^{2} - \Vs^{2})g_1+ {\laexp}_1(\Vs-\Vs^*) 
- \ii {\laexp}_1^2 f_1 + {\laexp}_2 V_s,\\
\lefteqn{  \sdd{\gexp}{\xi} 
+  \sigma |\Vs|^{2\sigma-2}(\Vs^*)^{2}  \fexp
+ ( \sigma+1)|\Vs|^{2 \sigma} \gexp
-  \gexp+\frac{\ii (\sigma-2) G}{2 \sigma} \gexp
+ \frac{G^2 \xi^2 }{4} \gexp}\hspace{2cm} &&\\
&=&   {\laexp}_1 \sigma |\Vs|^{2\sigma-2}(\Vs^2-(\Vs^*)^2)f_1
+{\laexp}_1  (\Vs^* - \Vs)
+ \ii {\laexp}_1^2 g_1 +  {\laexp}_2 V_s^*,
\eeqas
where we have neglected triply-exponentially-small terms involving ${\laexp}_2 {\laexp}_1$.
Multiplying by $\ii \Vs$, $-\ii\Vs^*$, adding and integrating the LHS is triply exponentially small. After simplifying, and neglecting the triply-exponentially-small term $(\Vs^2-(\Vs^*)^2){\laexp}_2$, the RHS gives
\begin{multline}
 \int_{-\infty}^\infty
 \ii  {\laexp}_1\sigma  |\Vs|^{2\sigma-2}((\Vs^*)^{2} - \Vs^{2})(\Vs g_1+\Vs^* f_1)\\+
 \ii {\laexp}_1(\Vs^2  -(\Vs^*)^2 )
+ {\laexp}_1^2 ( \Vs f_1 + \Vs^* g_1 )
\, \d \xi \sim 0. \label{solvla0}
\end{multline}
This is the equation which will determine the eigenvalue ${\laexp}_1$; note that it is quadratic, and ${\laexp}_1=0$ is a solution as expected.
In the outer region  $f_1$ and $g_1$ are exponentially small. Thus the integrals involving $f_1$ and $g_1$ are dominated by the inner region.
In the inner region
$f_1 = g_1 + $ exponentially small terms, and 
\[ \sdd{ f_1}{\xi} 
+5\Vs^{4}  f_1 
- f_1
+ \frac{G^2 \xi^2 }{4} f_1 =   \Vs.\]
We find
\beqas
f_1 & = & \frac{1}{2}\left( \frac{\Vs}{2} + \xi \fdd{\Vs}{\xi}\right) + G^2 f_{12},
\eeqas
where, up to exponentially small terms, 
\beqas
\sdd{ f_{12}}{\xi} +
5\Vs^{4}  f_{12} 
- f_{12}+ \frac{G^2 \xi^2 }{4} f_{12} &=&   \frac{1}{2} \xi^2 \Vs.
\eeqas
Unfortunately we need to find this correction term $f_{12}$ because the leading-order term will integrate to zero.
Expanding in powers of $G$, we find $f_{12} \sim- 2 V_1$ where $\Vs \sim V_0 + G^2 V_1 + \cdots$ (see \cite{jon1}), 
so that
\beqas
\int_{-\infty}^\infty (\Vs f_1 + \Vs^* g_1) \, \d \xi &\sim&
\int_{-\infty}^\infty \Vs  \left( \frac{\Vs}{2} + \xi \fdd{\Vs}{\xi} -4G^2 V_1\right)  \, \d \xi \sim  -4G^2\int_{-\infty}^\infty V_0 V_1  \, \d \xi = -\frac{G^2\sqrt{3} \pi^3}{64}.
\eeqas
The dominant contribution to the integral of $(\Vs^2 - (\Vs^*)^2)$ comes from the outer region before the turning point,
in which, with $\xi = \rho/G$, 
\[
\Vs \sim \frac{2^{1/2}a_0}{(4-\rho^2)^{1/4}}( \ee^{\phi(\rho)/G} + \gamma \ee^{-\phi(\rho)/G} ), \qquad \phi = - \int_0^\rho \left(1- \frac{\bar{\rho}^2}{4}\right)^{1/2}\, \d \bar{\rho},\qquad \gamma = \frac{\ii \ee^{-\pi/G}}{2},
\]
(see \cite{jon1}), so that
\beqas
\ii \int_\infty^{\infty} (\Vs^2 - (\Vs^*)^2)\, \d \xi & \sim &
\frac{2\ii}{G} \int_0^{2} (\Vs +\Vs^*)(\Vs -\Vs^*)  \, \d \rho
\sim \frac{2\ii}{G} \int_{0}^{2} 2\frac{2^{1/2} a_0\ee^{\phi(\rho)/G}}{(4 - \rho^2)^{1/4}}2 \frac{2^{1/2} a_0\gamma\ee^{-\phi(\rho)/G}}{(4 - \rho^2)^{1/4}}  \, \d \rho\\
& \sim &-\frac{16 \sqrt{3} \ee^{-\pi/G}}{G} \int_{0}^{2} \frac{   \d \rho}{(4 - \rho^2)^{1/2}}
=  -\frac{8 \sqrt{3} \pi\,  \ee^{-\pi/G}}{G}.
\eeqas
The final term in  Eq.~(\ref{solvla0}) is subdominant, so that, to leading order, 
Eq.~(\ref{solvla0}) gives
\beqa
{\laexp}_1 & = &- \frac{8 \sqrt{3}\, \pi \ee^{-\pi/G}}{G} \frac{64}{G^2\sqrt{3} \pi^3} =  -\frac{512\,\ee^{-\pi/G}}{G^3\pi^2}.\label{asy0}
\eeqa
In appendix \ref{reducedappendix} we show that the asymptotic behaviour (\ref{asy0}) can be determined much more simply from the reduced system derived in \cite{jon1}, which describes the slow evolution of $G$ in the vicinity of the bifurcation.

\begin{figure}
 \begin{center}
\begin{overpic}[width=0.8\textwidth]{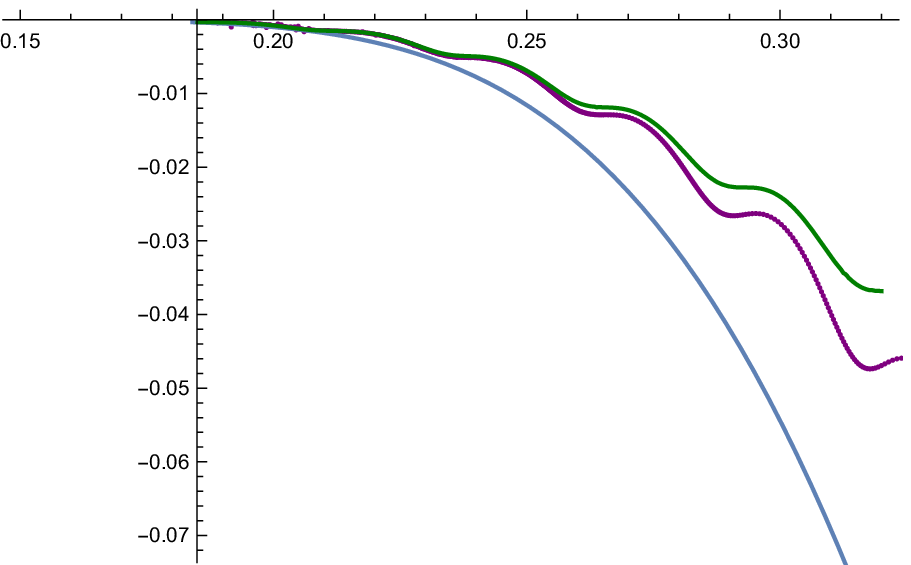}
\put(-5,30){\rotatebox{90}{$\lambda$}}
 \put(90,-7){$G$}
\end{overpic}\vspace{6mm}
    \end{center}
  \caption{Asymptotic prediction [cf. Eq.~(\ref{asy0})] (blue) compared to numerical solution (purple), for the eigenvalue in the vicinity of $\las=0$, for $K=20$. Also shown (green) is the approximation (\ref{la0num}).}
 \label{fig0asy}
\end{figure}

Figure \ref{fig0asy} shows the asymptotic prediction Eq.~(\ref{asy0}) against a direct numerical simulation. For this eigenvalue the convergence is slower as $G\ra0$ so that the leading-order approximation is not as close to the numerical solution. This is because the higher-order corrections are significant when estimating the integrals in Eq.~(\ref{solvla0}). To demonstrate this we also show in Fig.~\ref{fig0asy} the approximation
\beq {\laexp}_1 \sim -\ii\frac{ \int_{-\infty}^\infty (\Vs-(\Vs^*)^2)\, \d \xi}{ \int_{-\infty}^\infty (\Vs f_1-\Vs^*g_1)\, \d \xi}\label{la0num}
\eeq
with a numerical solution for $\Vs$, $f_1$ and $g_1$, which converges more quickly to the numerical value. It is clear that the latter expression
of Eq.~(\ref{la0num}) captures the dependence on $G$ more accurately
than the leading-order correction of the former.

\subsection{Continuous spectrum}
In the far field with $\rho = G \xi$, neglecting the exponentially small terms $|\Vs|^{2 \sigma}$ and $(\sigma-2)$,  we have
\beq
\ii \la f+ G^2\sdd{ f}{\rho} 
- f
+ \frac{\rho^2 }{4} f = 0,\qquad
- \ii \la g
+ G^2\sdd{g}{\rho} 
-  g 
+ \frac{\rho^2 }{4} g  = 0,\label{fgouter}
\eeq
along with the boundary conditions
\[
 G\fdd{f}{\rho} = \frac{\ii \rho f}{2}, \qquad
 G\fdd{g}{\rho} = -\frac{\ii \rho g}{2} \qquad \mbox{ at }\rho = \pm KG.
 \]
Let us start by imagining that these equations hold throughout the region $[-KG,KG]$, before returning to investigate the impact of the inner region near $\rho=0$.

Since the equations for $f$ and $g$ decouple (because we have ignored
the inner region), we can treat them separately, and each will give a
set of eigenvalues. In fact, we see that for any
eigenfunction-eigenvalue pair $(f,\la)$ the conjugates $(f^*,\la^*)$
satisfy the equations and boundary conditions for $g$.
We therefore start by focusing on the equation for $f$.

Using  the WKB expansion
\beq f =  A_f \ee^{\ii \phi_f/G} + B_f \ee^{-\ii \phi_f/G}, \label{contWKB}
\eeq
where
\[ A_f = \sum_{n=0}^\infty A_{fn}(\rho) (\ii G)^n, \qquad
B_f = \sum_{n=0}^\infty B_{fn}(\rho) (-\ii G)^n,\]
\[ \ii \la  -(\phi_f')^2 - 1 + \frac{\rho^2}{4} = 0,\]
 \[2 \ii \phi_f' A_{f0}' + \ii \phi_f'' A_{f0} = 0, \qquad
 - 2 \ii \phi_f' B_{f0}' - \ii \phi_f'' B_{f0}= 0,\]
we find 
 \[
 \phi_f = \int_0^{\rho} \left(\frac{\bar{\rho}^2}{4} - 1 + \ii \la \right)^{1/2} \, \d \bar{\rho}
 =  \frac{\rho }{4}\sqrt{\rho^2 - 4 + 4 \ii \la} + (-1 + \ii \la) \log\left(
 \frac{
 \rho + \sqrt{\rho^2 - 4 + 4 \ii \la}}{\sqrt{ - 4 + 4 \ii \la}}\right)
 ,\]
 with
 \[  A_{f0} = a_f\left(\frac{{\rho}^2}{4} - 1 + \ii \la \right)^{-1/4},\qquad
B_{f0} = b_f\left(\frac{{\rho}^2}{4} - 1 + \ii \la \right)^{-1/4}.
\]
Note that this expansion differs from that performed previously in that we have included $\la$ at leading order rather than assuming that $\la = O(G)$.

\begin{figure}
   \begin{center}
 \begin{overpic}[width=0.4\textwidth]{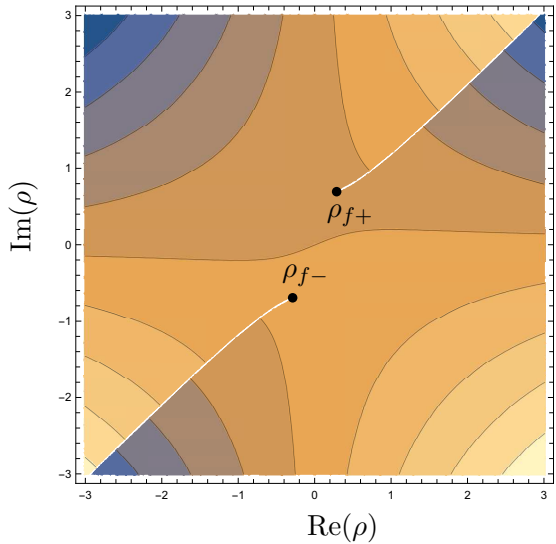}
\put(-10,50){\rotatebox{90}{$\im(\rho)$}}
\put(50,-7){$\re(\rho)$}
\put(54,57){$\rho_{f+}$}
\put(45,45){$\rho_{f-}$}
  \end{overpic}\vspace{6mm}
    \end{center}
  \caption{Branch points and branch cuts in $\phi_f$ when $\la=-0.1-1.1 \ii$. The contour shading corresponds to $\im(\phi_f)$.}
 \label{figbranches}
\end{figure}

 There are two turning points, at
 \[ \rho = \rho_{f\pm} = \pm 2(1-\ii \la)^{1/2}.\]
 In order to define uniquely $\phi_f$ let us put branch cuts from these turning points to $\pm \ii \infty$ away from the real axis, as indicated in Fig. \ref{figbranches} for an arbitrary but representative value of $\la$.


  \begin{figure}
   \begin{center}
\subcaptionbox{$\la=-0.3 - 0.5 \ii$}{\begin{overpic}[width=0.46\textwidth]{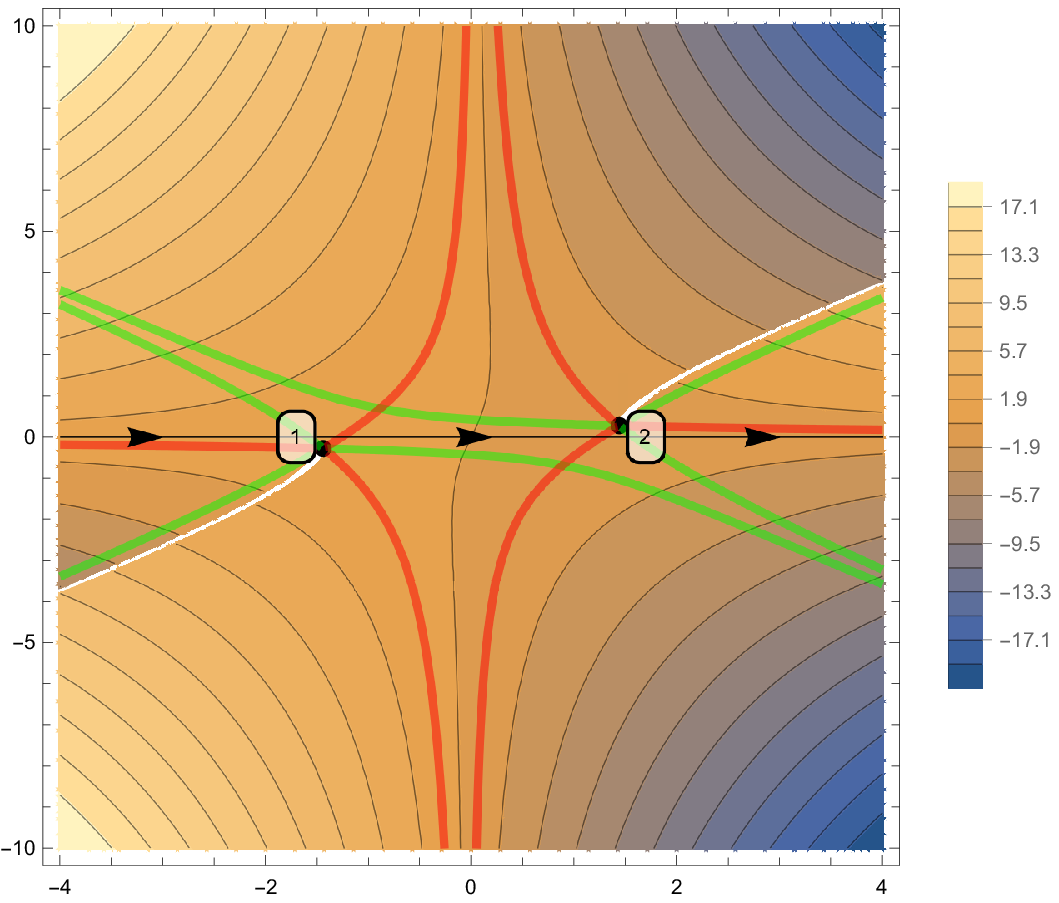}
\put(58,48){{\small $\rho_{f+}$}}
\put(30,38){{\small $\rho_{f-}$}}
\end{overpic}
  \vspace{6mm}}
\subcaptionbox{$\la=-0.3 - 2 \ii$}{\begin{overpic}[width=0.46\textwidth]{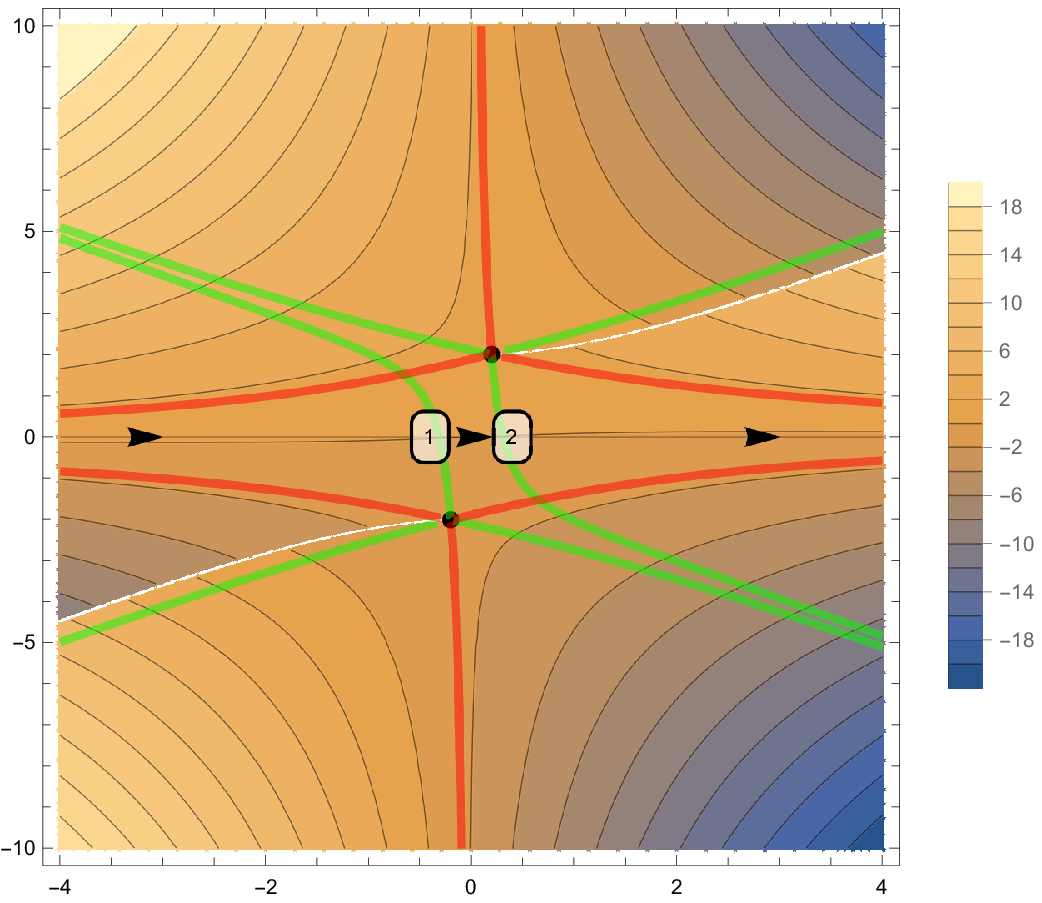}
\put(48,55){{\small $\rho_{f+}$}}
\put(40,32){{\small $\rho_{f-}$}}
  \end{overpic}
  \vspace{6mm}}
    \end{center}
  \caption{Stokes lines (green) and anti-Stokes lines (red). The contour shading corresponds to $\im(\phi_f)$. The path along the real axis is indicated, as well as the two points at which Stokes lines are crossed.}
 \label{figStokes}
\end{figure}
To impose the boundary conditions on Eq.~(\ref{contWKB}) we need to take account of the change in the coefficients $a_f$ and $b_f$ due to Stokes phenomenon
(see e.g.~\cite{MullerKirsten, chapman}). In Fig.~\ref{figStokes} we illustrate the Stokes lines associated with each of the turning points for various values of $\la$.
 Let us calculate the change in the coefficients $a_f$ and $b_f$ as we cross Stokes lines when moving from $\rho=-KG$ to $\rho = KG$.
 We suppose that $\re(\la)<0$ so that $\rho_+$ is in the first quadrant.
 Although the topology of the anti-Stokes lines changes as $\im(\la)$ varies,  as we move along the real axis from minus infinity to infinity we always cross one Stokes line from each turning point. Across these Stokes lines the dominant WKB approximation will turn on a multiple of the subdominant WKB approximation.

\begin{figure}
   \begin{center}
 \begin{overpic}[width=0.4\textwidth]{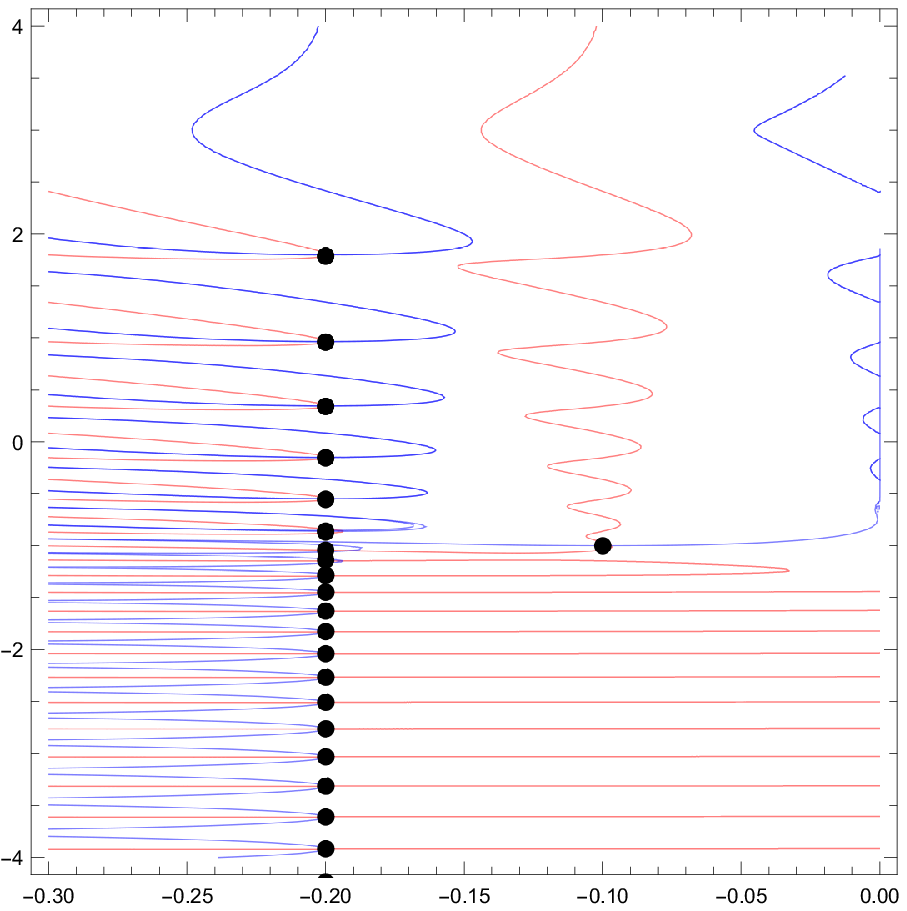}
\put(-10,50){\rotatebox{90}{$\im(\la)$}}
 \put(50,-7){$\re(\la)$}
  \end{overpic}\vspace{6mm}
    \end{center}
  \caption{Eigenvalues for the problem of Eq.~(\ref{fgouter}) for $f$ (black
    points), for $G=0.2$. The contours show the asymptotic prediction of Eq.~(\ref{evcontout}). The
    red curves correspond to $\im(Q_f - \ii \ee^{(1-\ii\la)\pi/G}) =
    0$, while the blue curves correspond to $\re(Q_f - \ii
    \ee^{(1-\ii\la)\pi/G}) \in \{-1,1\}$. The eigenvalues should lie at
    the intersections of these contours. The approximation is very
    good, apart from near $\la=-\ii$, at which point the two turning
    points $\rho_{f\pm} = \pm 2(1-\ii\la)^{1/2}$ coalesce.}
 \label{figeigsouter}
\end{figure}

 The first Stokes line we cross, indicated by a ``1'' in Fig.~\ref{figStokes}, is that which moves up from $\rho_{f-}$, on which 
 $\im(\phi_f(\rho))>\im(\phi_f(\rho_{f-}))$. Thus 
 $\ee^{-\ii \phi_f/G}$ is the exponentially dominant term, and the coefficient $a_f$ changes by $-\ii b_f \ee^{-2\ii \phi_f(\rho_{f-})/G} $.
 The second Stokes line we cross, indicated by a ``2''  in Fig.~\ref{figStokes}, is that which moves down from $\rho_{f+}$, on which $\im(\phi_f(\rho))<\im(\phi_f(\rho_{f+}))$. This on this Stokes line $\ee^{\ii \phi_f/G}$ is the exponentially dominant term, and the coefficient $b_f$ changes by $\ii a_f \ee^{2\ii \phi_f(\rho_{f+})/G}$.
Thus, together, the change in the coefficient is
 \beqas
 (a_f^{-\infty},b_f^{-\infty}) &\ra&  (a_f^{-\infty}-\ii b_f^{-\infty} \ee^{-2\ii \phi_f(\rho_{f-})/G} ,b_f^{-\infty})\\
 &\ra& (a_f^{-\infty}-\ii b_f^{-\infty} \ee^{-2\ii \phi_f(\rho_{f-})/G} ,b_f^{-\infty} + \ii (a_f^{-\infty}-\ii b_f^{-\infty} \ee^{-2\ii \phi_f(\rho_{f-})/G})\ee^{2\ii \phi_f(\rho_{f+})/G}),
 \eeqas
so that 
 \beqas
 a_f^{\infty} & = & a_f^{-\infty}-\ii b_f^{-\infty} \ee^{-2\ii \phi_f(\rho_{f-})/G},\\
 b_f^{\infty} & = & b_f^{-\infty} + \ii a_f^{-\infty}\ee^{2\ii \phi_f(\rho_{f+})/G}+ b_f^{-\infty} \ee^{-2\ii \phi_f(\rho_{f-})/G}\ee^{2\ii \phi_f(\rho_{f+})/G}.
 \eeqas
 The boundary condition at $\rho = KG$ gives, at leading order,
\beqas
\ee^{2 \ii \phi_f(KG)/G} \frac{\sqrt{(KG)^2 - 4 + 4 \ii \la}  - KG
  }{\sqrt{(KG)^2 - 4 + 4 \ii \la}+ KG}&=&
\frac{B_{f0}(KG)}{A_{f0}(KG)} = \frac{b_f^{\infty}}{a_f^{\infty}}.
\eeqas
The boundary condition at $\rho = -KG$ gives, at leading order,
\beqas
\ee^{2 \ii \phi_f(-KG)/G} \frac{\sqrt{(KG)^2 - 4 + 4 \ii \la}  + KG
  }{\sqrt{(KG)^2 - 4 + 4 \ii \la}- KG}&=&
\frac{B_{f0}(-KG)}{A_{f0}(-KG)} =
\frac{b_f^{-\infty}}{a_f^{-\infty}} .
\eeqas
Noting that  $\phi_f$ is odd, if we let
\begin{eqnarray}
Q_f = \ee^{2 \ii \phi_f(KG)/G} \frac{\sqrt{(KG)^2 - 4 + 4 \ii \la}  - KG
  }{\sqrt{(KG)^2 - 4 + 4 \ii \la}+ KG},
  \label{extra1}
  \end{eqnarray}
then we have the following homogeneous system of four equations in the four
unknowns $a_f^{\infty}$, $b_f^{\infty}$, $a_f^{-\infty}$, $b_f^{-\infty}$,
\beqas
b_f^{\infty} &=& Q_f a_f^{\infty},\\
a_f^{-\infty} &=& Q_f b_f^{-\infty},\\
 a_f^{\infty} & = & a_f^{-\infty}-\ii b_f^{-\infty} \ee^{-2\ii \phi_f(\rho_{f-})/G},\\
 b_f^{\infty} & = & b_f^{-\infty} + \ii a_f^{-\infty}\ee^{2\ii \phi_f(\rho_{f+})/G}+ b_f^{-\infty} \ee^{-2\ii \phi_f(\rho_{f-})/G}\ee^{2\ii \phi_f(\rho_{f+})/G}.
 \eeqas
 Noting that $\phi_f(\rho_{f+}) = - \phi_f(\rho_{f-}) = -(\la+\ii)\pi/2$, the condition for a non-trivial solution is
  \beq (Q_f - \ii\ee^{(1-\ii\la)\pi/G})^2 = 1.\label{evcontout}
  \eeq
  For finite $K$, Eq.~(\ref{evcontout}) gives a discrete set of eigenvalues with the separation between neighbouring eigenvalues becoming smaller as $K\ra \infty$, approximating the continuous spectrum.
For large $K$, 
 \[ Q_f \sim  \ee^{\ii K^2G/2} \ee^{-\ii(1+ \ii \la)/G} (-1+\ii \la)^{1+(\ii + \la)/G} (KG)^{-2-2(\ii + \la)/G}.\]
 For $\im(\la)<0$ we need $|Q_f| \sim 1$ as $G \ra 0$ to get a
 balance in Eq.~(\ref{evcontout}), which requires $\re(\la) \sim -G$.
For $\im(\la)>0$ we need $Q_f \sim \ii\ee^{(1-\ii\la)\pi/G}$ as $G \ra 0$ to get a
balance in Eq.~(\ref{evcontout}), which also requires
$\re(\la) \sim -G$. Thus the continuous spectrum of the problem [cf. Eq.~(\ref{fgouter})] lies close to (but not exactly on) the line $\re(\la) = -G$.
The calculation for $g$ is similar
and gives the same equation as Eq.~(\ref{evcontout}) with $\la \ra \la^*$
as expected. In Fig.~\ref{figeigsouter} we compare the predictions of Eq.~(\ref{evcontout}) (recalling the definition of Eq.~(\ref{extra1})) 
with the numerical evaluation of the eigenvalues of  Eq.~(\ref{fgouter}). A very good agreement is found between the latter
(identified as black dots) and the former (identified via the
intersection of the contours of the blue and red curves associated
with the real and the imaginary parts of Eq.~(\ref{evcontout}).

We now consider how the picture above changes when we include the
inner region. Then, in addition to the Stokes lines already
considered, there is a change in the coefficients $a$ and $b$ as we
pass from $\rho=0-$ to $\rho=0+$. The connection formula comes from matching the solution in
the inner region with the far field expansions on each side.
Note  that $f$ and $g$ are coupled in the inner region, so that we
no longer have two separate eigenvalue problems. Specifically, at
leading order in the inner region $\Vs$ is real and 
\beqa
\ii\la f+\sdd{ f}{\xi} +
2 \Vs^{4} g  
+ 3\Vs^{4}  f 
- f
&=& 0,\label{in1}\\
-\ii  \la g
+ \sdd{g}{\xi} 
+  2\Vs^{4} f 
+ 3\Vs^{4} g 
-  g 
  &=& 0,\label{in2}
\eeqa
with
\beqas
f & \sim & a_f^{0+}\frac{\ee^{\ii \phi_f(0)/G}}{(-1+\ii \la)^{1/4}}\ee^{(1- \ii \la)^{1/2}\xi}+ b_f^{0+}\frac{\ee^{\ii \phi_f(0)/G}}{(-1+\ii \la)^{1/4}}\ee^{-(1- \ii \la)^{1/2}\xi} \mbox{ as } \xi \ra \infty,\\
g & \sim &a_g^{0+}
\frac{\ee^{\ii \phi_g(0)/G}}{(-1-\ii \la)^{1/4}}
\ee^{-(1+\ii \la)^{1/2}\xi}+ b_g^{0+}
\frac{\ee^{\ii \phi_g(0)/G}}{(-1-\ii \la)^{1/4}}
\ee^{(1+ \ii \la)^{1/2}\xi}  \mbox{ as } \xi \ra \infty,\\
f & \sim & a_f^{0-}\frac{\ee^{\ii \phi_f(0)/G}}{(-1+\ii \la)^{1/4}}
\ee^{(1- \ii \la)^{1/2}\xi}+ b_f^{0-}
\frac{\ee^{\ii \phi_f(0)/G}}{(-1+\ii \la)^{1/4}}\ee^{-(1- \ii \la)^{1/2}\xi}  \mbox{ as } \xi \ra -\infty,\\
g & \sim & a_g^{0-}
\frac{\ee^{\ii \phi_g(0)/G}}{(-1-\ii \la)^{1/4}}
\ee^{-(1+ \ii \la)^{1/2}\xi}+
b_g^{0-}
\frac{\ee^{\ii \phi_g(0)/G}}{(-1-\ii \la)^{1/4}}
\ee^{(1+ \ii \la)^{1/2}\xi}  \mbox{ as } \xi \ra -\infty.
\eeqas
The solution gives connection formulae between the incoming coefficients  
$a_f^{0-}$, $b_f^{0-}$, $a_g^{0-}$ and $b_g^{0-}$ and the outgoing
coefficients   $a_f^{0+}$,
$b_f^{0+}$, $a_g^{0+}$ and  $b_g^{0+}$. Unfortunately it is not
possible to determine these formulae analytically.

However, when $\la$ is large the first two terms in Eqs.~(\ref{in1}) and (\ref{in2}) dominate, and the solution is of WKB form even in the inner
region. The phase factor is trivial, with no turning points, so that there is no
change in coefficient. Thus for large $\la$ the eigenvalues should be well approximated by Eq.~(\ref{evcontout}).

On the other hand when $\la=0$ we find by solving  Eqs.~(\ref{in1})-(\ref{in2}) as a power series in $G$ that
\[
  a_f^{0+} \sim  a_g^{0-}, \quad
  b_f^{0+} \sim -b_g^{0-}, \quad
  b_g^{0+} \sim b_f^{0-}, \quad
  a_g^{0+} \sim - a_f^{0-},
\]
so that there must be some mixing of the coefficients in $f$ and $g$
for small $\la$. Identifying the details of the relevant spectrum
at small $\lambda$ remains a challenging question for future study.

\section{Conclusions \& Future Challenges}\label{conclusions}

In the present work, we have revisited the topic of stability
of solutions that are self-similarly blowing up. We followed up on the
earlier work of~\cite{siettos} with substantially improved
numerical means and techniques, and also added a systematic
theoretical understanding, building also on important works
in the intermediate time interval (such as the key
contributions of~\cite{wit1,wit2}).
This has allowed us to obtain a systematic understanding of the
3 eigenvalue pairs of the Hamiltonian system at the critical
point of $\sigma d=2$ and its continuous spectrum. We advocated
the relevance of exploring the self-similar solutions in the
co-exploding frame, by analogy with the study of traveling solutions
in a co-traveling frame, as per the standard dynamical systems
perspective~\cite{bjorn,promislow}. We have also explained 
systematically why, despite the presence of positive real 
eigenvalues, the relevant self-similar solution is not genuinely
unstable but only subject to effectively neutral eigendirections.
To corroborate the relevant results, we performed
direct numerical simulations in the renormalized frame,
verifying (in line with earlier computations) the attractivity 
of the relevant waveforms. 

Naturally, this analysis raises a number of interesting questions
for further research. Understanding the dynamics (and the potential
role of self-similarity) slightly below the critical point
$\sigma d=2$ is an example of this type. Moreover, we 
have argued that the supercritical solutions considered
herein are effectively stable, upon explaining the origin
of their real eigendirections. Yet, it is well-known that there
are other problems for which multiple, higher-order collapsing
solutions branches exist, some among which are dynamically
unstable: a notable example of this sort is, e.g., the complex
Ginzburg-Landau equation~\cite{vivi}. It is then of particular
interest to explore such waveforms via the type of spectral
analysis proposed herein, and corroborate in a systematic
fashion their stability or instability, as well as leverage
such spectral information in an attempt to understand the corresponding
direct numerical simulations of the relevant system in the renormalized
frame. In a different vein, there are other important dispersive
PDE models that feature similar bifurcations towards the emergence
of collapsing solutions, such as the generalized KdV 
problem; for a recent exposition of the collapsing solutions
and asymptotics thereof, see, e.g.,~\cite{budd_recent}. It is
then of particular interest to adapt the methodology proposed
herein to the latter problem to explore the potential generality
of the eigenvalue phenomenology identified in the present work.
Such topics are presently under active investigation and relevant
results will be
reported in future publications.

\vspace{5mm}

{\it Acknowledgments.} This material is based upon work supported by the US
National Science Foundation under Grants No. DMS-1809074 and 
PHY-2110030 (P.G.K.). PGK and EGC are also grateful to Dionyssis
Mantzavinos for numerous useful discussions during the early stages
of the present work.

\appendix

\section{An Instructive ODE Example}

Consider, arguably, one of the simplest self-similar problems, 
namely the ODE:
\begin{eqnarray}
  \dot{x}=x^p, \quad x(t)\in\mathbb{R}.
  \label{seq1}
\end{eqnarray}
In order to leverage the self-similar frame to analyze this
problem, we seek to absorb the temporal dependence through a transformation
to go to a frame where the solution appears steady. In this (ODE) 
case, it is not a steady spatial profile, as in the PDE example studied
throughout this work, but instead a ``number.'' We thus use 
$x(t)=A(\tau) \bar{x}$ with $\tau=\tau(t)$ corresponding to a rescaling
of time (to be determined), and obtain the steady-state problem 
$\bar{x}=\bar{x}^p$ leading to $\bar{x}=0$ or $\bar{x}=1$. Then, 
in accordance to general self-similarity principles~\cite{baren,zhang2}, 
we select
\begin{eqnarray}
  \frac{A_{\tau}}{A}=1\Rightarrow A=A_0 e^{\tau}; \quad
  \tau_t=A^{p-1} \Rightarrow A(t)=\left[\frac{1}{(p-1)
  (t^{\star}-t)}\right]^{\frac{1}{p-1}} =A_0 \left[
  \frac{1}{(t^{\star}-t)}\right]^{\frac{1}{p-1}}.
  \label{seq2}
\end{eqnarray}
In Eq.~\eqref{seq2}, $t^{\star}$ denotes the blow-up time, i.e.,
$x(t)\to \infty$ as $t\nearrow t^{\star}$. In the self-similar 
frame, we have indeed devised a much more elaborate way to obtain a 
simple ODE result. The innate advantage of the method, however,
is that in this frame that ``explodes'' with the solution it is
possible to perform a stability analysis using:
\begin{eqnarray}
x(t)=A(\tau) \left[\bar{x} + \epsilon y(\tau)\right] \Rightarrow
y_{\tau}=(p-1) y, \quad \epsilon\ll 1,
\label{seq3}
\end{eqnarray}
assuming that we keep the leading order (O$(\epsilon)$) terms in 
$y$. We thus observe that the self-similar frame features a single
eigenvalue (indeed, since it is an ODE rather than a PDE) of 
$\lambda=p-1$. For collapsing solutions with $p>1$, this is an 
eigenvalue associated with growth since $\lambda>0$.

A natural question then is whether this is a {\it true} instability.
The perturbation $A(\tau) \epsilon y$ can be rewritten as
$\epsilon e^{p \tau} \sim \epsilon (t^{\star}-t)^{p/(1-p)}$.
But then, considering a shift in the collapse 
time $t^{\star}\rightarrow t^{\star} + \tilde{\epsilon} \delta t$
($\tilde{\epsilon}\ll 1$), and substituting it in the 
original solution, we obtain:
\begin{eqnarray}
x(t)\to %
\widetilde{x}(t)=\left[\frac{1}{p-1}\right]^{\frac{1}{p-1}} \left[\frac{1}{t^{\star}+
  \tilde{\epsilon} \delta t -t}\right]^{\frac{1}{p-1}}
  =A_0 \left[\frac{1}{t^{\star}-t}\right]^{\frac{1}{p-1}}
  - \left(\frac{\tilde{\epsilon} A_0 \delta t}{p-1} \right)
  (t^{\star}-t)^{\frac{p}{1-p}}.
  \label{seq4}
\end{eqnarray}
Namely, the positive eigenvalue does {\it not} correspond to a
true instability but rather is associated with the translational
invariance of the ODE with respect to the shifting of the
collapse time. 

\section{Continuous spectra obtained from the rescaled NLS equation  }\label{sec:large}
\renewcommand{\thefigure}{A\arabic{figure}}
\setcounter{figure}{0}

\begin{figure}
\begin{center}
\begin{overpic}[width=0.95\textwidth]{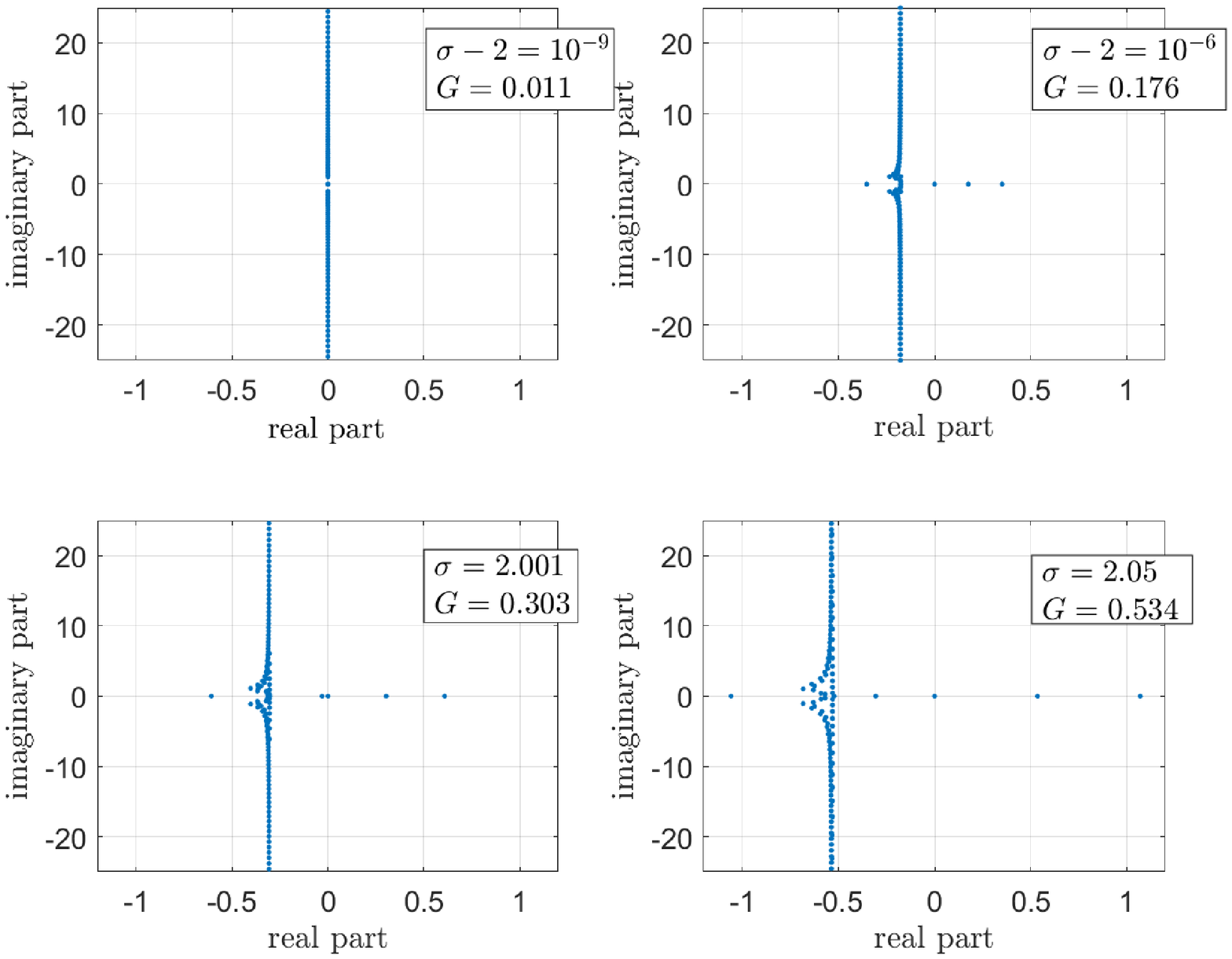}
\end{overpic}
\end{center}
  \caption{Continuous spectra obtained from the numerical solution of the rescaled NLS equation with $K=20$ and $\sigma$ values close to the critical value, $\sigma=2$: $\sigma=2+10^{-9}$ (top left panel) and $\sigma=2+10^{-6}$ (top right panel), as well as $\sigma=2.001$ (bottom left panel), $\sigma=2.05$ (bottom right panel).}
  \label{figAppendix1}
\end{figure}

Fig.~\ref{figAppendix1} depicts numerical spectra of the rescaled NLS equation obtained for different $\sigma$ values. One can  observe the alignment of eigenvalues on a nearly vertical line of continuous spectrum, with real part, $\lambda_r=-G$. 
As we move away from the critical value, $\sigma=2$, the distortions from the vertical line become progressively larger. 

\section{Effect of size domain on the numerical computations }\label{sec:Keffect}
\begin{figure}
\begin{center}
\begin{overpic}[width=0.95\textwidth]{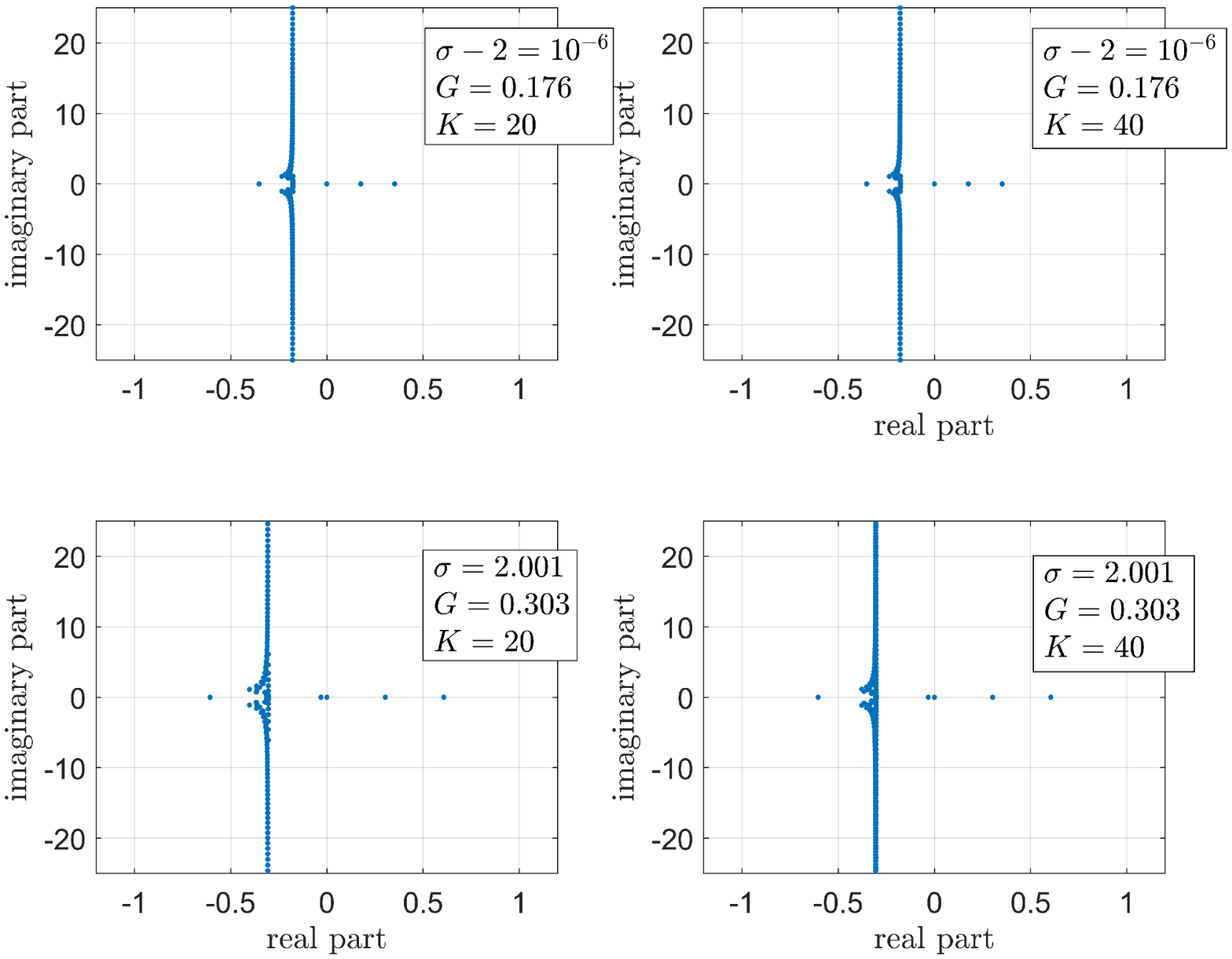}
\end{overpic}
\end{center}
  \caption{Comparison of spectra obtained from the numerical solution of the rescaled NLS equation for size domain $K=20$ (left panel) and $K=40$ (right panel)}
  \label{figAppendix2}
\end{figure}

Fig.~\ref{figAppendix2} presents a comparison of the computed spectra of the rescaled NLS equation for size domain $K=20$ and $K=40$. Even by doubling the size of the computational domain, the alignment of eigenvalues on a ``vertical'' line of the continuous spectrum remains practically unchanged, and the real eigenvalues deviations are also visually indistinguishable.

\section{Stability of the reduced system}
\label{reducedappendix}
In \cite{jon1}, asymptotic solutions of Eq.~(\ref{pde9}) were constructed which
result in the following reduced system for the growth rate $G$ and an associated variable $\beta$, which can be thought of as the  normal form associated with the bifurcation:
\beqa
\fdd{G}{\tau} & = & \beta - G^2,\label{1}\\
c_0 \fdd{\beta}{\tau} & = & \frac{(\sigma-2)}{2 \sigma} b_0 G - A_{1}^2 \sign(G) \ee^{-\pi/|G|},\label{2}
\eeqa
where
\[ c_0 = \frac{\sqrt{3}\, \pi^3}{512}, \qquad b_0 = \frac{\sqrt{3}\, \pi}{4},\qquad A_1 = 12^{1/4}.\]

\subsection{Stability of the zero solution}
Linearizing about the origin gives
\beqas
\fdd{G}{\tau} & = & \beta ,\\
 \fdd{\beta}{\tau} & = & \frac{(\sigma-2)b_0}{2 \sigma c_0}  G.
\eeqas
The characteristic equation is
\[ \la^2 = \frac{(\sigma-2)b_0}{2 \sigma c_0},\]
showing a pair of eigenvalues moving from the imaginary axis to the real axis as $\sigma$ passes through 2 as expected.

\subsection{Stability of the non-zero solution}
The  steady solution ($G_0,\beta_0$) given by
\[ \beta_0 = G_0^2, \qquad  \frac{(\sigma-2)}{2 \sigma} b_0 G_0 = A_{1}^2  \ee^{-\pi/G_0}.
\]
with $G_0>0$ corresponds to the blow-up solution in the original frame. Perturbing about this solution by writing $G = G_0+x$, $\beta = G_0^2+y$ and linearizing gives
\beqas
\fdd{x}{\tau} & = & y - 2 G_0 x,\\
c_0 \fdd{y}{\tau} & = & \frac{(\sigma-2)}{2 \sigma} b_0 x - A_{1}^2 \ee^{-\pi/G_0}\frac{\pi}{G_0^2}x.
\eeqas
The characteristic equation is
\[ \la(\la+2G_0) - \frac{1}{c_0}\left( \frac{( \sigma-2)}{2 \sigma} b_0  - A_{1}^2 \ee^{-\pi/G_0}\frac{\pi}{G_0^2}\right) = 0.
\]
One eigenvalue is exponentially close to zero, while one is exponentially close to $-2G_0$.
The exponentially small  eigenvalue is approximately
\beqas
\la & \sim & \frac{1}{2c_0 G_0}\left( \frac{( \sigma-2)}{2 \sigma} b_0  - A_{1}^2 \ee^{-\pi/G_0}\frac{\pi}{G_0^2}\right)
 \sim  - A_{1}^2 \ee^{-\pi/G_0}\frac{\pi}{2c_0 G_0^3}
 \sim - \frac{\ee^{-\pi/G_0}}{512 G_0^3 \pi^2},
\eeqas
in agreement with the detailed calculation of Section \ref{sec:zeroev}. Note that this is also the eigenvalue which remains when reducing Eqs.~(\ref{1})-(\ref{2}) to the slow manifold
giving
\beqa
2 c_0 G \fdd{G}{\tau} & = & \frac{( \sigma-2)}{2 \sigma} b_0 G - A_{1}^2 \sign(G) \ee^{-\pi/|G|}.\label{3}
\eeqa
We note that the exponentially small correction to the eigenvalue $\la \sim -2G$ does not agree with the detailed calculation in Section \ref{sec:laM2G}. This is due to the fact that it corresponds to an exponentially small correction of the fast timescale, and the system of Eqs.~(\ref{1})-(\ref{2}) is accurate at leading-order only in that regime.

\end{document}